\documentclass[]{jkpapersmallfont}
\usepackage{tikz}
\usepackage{aas_macros}
\usetikzlibrary{decorations.markings}

\title{Holography from\texorpdfstring{\\}{ }Decoherence and Entanglement}
\author{Josh Kirklin}
\email{jjvk2@cam.ac.uk}
\institution{Department of Applied Mathematics and Theoretical Physics, Centre for Mathematical Sciences, University of Cambridge, Cambridge, UK%
}

\abstr{%
    We describe an explicit mechanism for the emergence of a dynamical holographic bulk from the structure of entanglement in a quantum state. We start with a generic system in complete isolation, assuming it has a classical limit involving coherent states. Then we entangle it with another system of that kind, and subject the pair to a decohering process. We make a number of broadly applicable and physically reasonable assumptions about this setup. First, we assume that the states selected by the decoherence (called pointer states) have the same local symmetries as the isolated systems, in a sense which is made precise. We also assume that the modular Hamiltonians of pointer states scale inversely with Planck's constant, so that the pointer states are highly entangled in the classical limit. Finally, we require the timescale of decoherence to scale in a certain way with Planck's constant, so that decoherence happens very frequently in the classical limit, but not too frequently. Given these assumptions, we demonstrate that the semiclassical evolution of the system is dominated by a certain dynamical generalisation of Uhlmann holonomy. We construct a coherent state path integral for this evolution, showing that the semiclassical fields evolve in a spacetime with one more dimension than the isolated case. The additional dimension is generated by modular flow.
}

\bibliography{refs}

\begin{document}

\maketitleandtoc

\section{Introduction}
\label{Section: Introduction}

It has become increasingly clear that there is a deep connection between entanglement and the structure of spacetime in quantum gravity. This idea has its roots in the realisation that the Bekenstein-Hawking entropy of a black hole~\cite{BekensteinEntropy,HawkingEntropy} can be attributed to entanglement between degrees of freedom on either side of the horizon~\cite{QuantumSourceOfEntropy,EntropyAndArea}; in holography this was generalised to the Hubeney-Rangamani-Ryu-Takayanagi (HRT) formula~\cite{RyuTakayanagi1,RyuTakayanagi2,HRT}, which associates the areas of a large class of bulk surfaces with entanglement entropies of appropriate subsystems. Many other similar relationships have been proposed, equating various measures of entanglement with other geometric properties of the bulk spacetime. Motivated additionally by the fact that an eternal black hole spacetime is holographically dual to a thermofield double~\cite{EternalBlackHoles} (which has a very specific pattern of entanglement), this led to the suggestion that these relationships are more than just a coincidence, and that the bulk spacetime itself somehow emerges from the entanglement in the quantum state~\cite{BuildingUpSpacetime,EREPR}. An important perspective on this has come from tensor network and quantum error correction approaches to holographic duality~\cite{Swingle1,Swingle2,QuantumErrorCorrection,ToyModels,BeyondToyModels}.

Despite all this, an explanation of how exactly the bulk spacetime emerges has been lacking. Additionally, it is not at all clear what role entanglement plays in the string theoretic arguments which underpin the most concrete example of holography, AdS/CFT~\cite{AdSCFTMaldacena,AdSCFTWitten}. Indeed, if gravity actually is a consequence of entanglement, then this shouldn't depend on any of the fine details of the fundamental theory -- only on whether it provides the right type of entanglement. So what is really needed is a more `phenomenological' perspective, in which we make some basic qualitative assumptions about the entanglement and dynamics of the quantum theory, and then see if those assumptions lead to holography. The assumptions should be as broadly applicable as possible, and physically reasonable. In this paper, we discuss one possible approach from this point of view.

The starting point is a generic isolated quantum system with a classical limit $\hbar\to0$; this system should be thought of as part of the lower-dimensional `boundary theory' on one side of the holographic duality. The parameter $\hbar$ is usually Planck's constant -- it could also be $1/N$ in a large $N$ gauge theory, or something else, but for simplicity we will just continue to use the symbol $\hbar$. A particularly useful and general description of the classical limit is given by coherent states. We will give some relevant basic facts about this description; for more information, see for example~\cite{LargeNLimits,CoherentStates}. One fixes a Lie group $G$ consisting of all the possible dynamical transformations that can be performed on the physical system. For each value of $\hbar$ we assume there is a Hilbert space $\mathcal{H}=\mathcal{H}_\hbar$, and a unitary irreducible representation $u=u_\hbar$ of $G$ acting on $\mathcal{H}$. We also pick a normalised `base' state $\ket{0}\in\mathcal{H}$, and by acting on $\ket{0}$ with $u$ we then obtain a set of states
\begin{equation}
    \{\ket{x} = u(x)\ket{0},\quad x\in G\}.
\end{equation}
These are the coherent states. For the classical limit to exist, we require the Berry connection of these states
\begin{equation}
    i\mel{x}{\mathrm{d}}{x},
\end{equation}
which is a real 1-form on $G$, to be $\order{1/\hbar}$ as $\hbar\to 0$. Operators can depend on $\hbar$, so when we talk about an `operator', what we really mean is a family of operators, one acting on each Hilbert space $\mathcal{H}_\hbar$. The coherent states allow us to discuss the asymptotics of these operators as $\hbar\to 0$. In particular, when we write 
\begin{equation}
    O = \order{f(\hbar)}
\end{equation}
for some function $f(\hbar)$, what we mean is that the coherent state correlators of $O$ obey
\begin{equation}
    \frac{\mel{x_1}{O}{x_2}}{\braket{x_1}{x_2}} = \order{f(\hbar)} \quad \text{for all } x_1,x_2\in G.
\end{equation}
A special case of this is $O=\order{1}$; an operator with this property is called a `classical' operator. The second requirement for the existence of a classical limit is that the Hamiltonian $H$ is a classical operator. In~\cite{LargeNLimits}, it is shown that $\comm{O_1}{O_2}=\order{\hbar}$ for any two classical operators $O_1,O_2$, and a corollary of this is that
\begin{equation}
    e^{iO_1/\hbar}O_2e^{-iO_1/\hbar} = \order{1}.
    \label{Equation: classical automorphism}
\end{equation}
So the automorphism generated by $iO_1/\hbar$ preserves the asymptotics of any operator. Because $u$ is an irrep, Schur's lemma implies that the coherent states give a resolution of the identity
\begin{equation}
    I = \int \dd{x}\ket{x}\bra{x},
    \label{Equation: resolution}
\end{equation}
where $\dd{x}$ is the invariant measure on $G$ (appropriately normalised). By inserting this many times, we can write the transition amplitude after a time $T$ between coherent states $\ket{x}$ and $\ket{x'}$ as a path integral in the usual way, obtaining
\begin{equation}
    \mel{x'}{e^{-iH T/\hbar}}{x} = \int \Dd{x}\exp(iS[x]/\hbar),
\end{equation}
where the integral is done over paths $x(t)$ which begin at $x$ and end at $x'$, and the action is
\begin{equation}
    S[x] = \int_0^T\Big(i\hbar\braket{x}{\dot{x}} - \mel{x}{H}{x}\Big)\dd{t}.
    \label{Equation: isolated action}
\end{equation}
By the requirements on the Berry connection and Hamiltonian, we have $S[x] = \order{1}$, so this can be treated as a classical action, and we can apply the usual methods of stationary phase to the path integral.

In this paper, we will consider a pair of subsystems with classical limits in terms of coherent states, corresponding to two disconnected components of the `boundary theory'. We label the two subsystems $A,B$, and use $\mathcal{H}_A,\mathcal{H}_B$ to denote their respective Hilbert spaces. We will want to consider a limit in which these two systems are very highly entangled. To make this precise, we take inspiration from a known feature of AdS/CFT. Suppose the combined state of the two systems is $\ket{\psi}$, so that the reduced states in the two subsystems are given by
\begin{equation}
    \rho_A = \tr_B\ket{\psi}\bra{\psi},\qquad \rho_B = \tr_A\ket{\psi}\bra{\psi},
\end{equation}
where $\tr_A,\tr_B$ denote partial traces over $\mathcal{H}_A,\mathcal{H}_B$ respectively. We will assume that $\rho_A,\rho_B$ are invertible. The modular Hamiltonians of $\ket{\psi}$ in $A$ and $B$ are
\begin{equation}
    K_A = -\log\rho_A,\qquad K_B = -\log\rho_B.
\end{equation}
These operators contain a large amount of information about the entanglement between $A$ and $B$. For example, the entanglement entropy may be computed with 
\begin{equation}
    S_A = \ev{K_A}_{\rho_A} = \tr(\rho_AK_A).
\end{equation}
In AdS/CFT, the modular Hamiltonian of a boundary subregion is given at leading order by~\cite{BulkRelativeEntropy}
\begin{equation}
    K = \frac{\hat{A}}{4 G_{\text{N}} \hbar}+\dots.
\end{equation}
Here $\hat{A}$ is an operator which gives the area of the HRT surface corresponding to the boundary subregion, and $G_{\text{N}}$ is Newton's constant. For us, the important feature of this formula is the factor of $1/\hbar$, which implies that there is a very large amount of entanglement in the classical limit $\hbar\to 0$. We will assume that this scaling holds for the modular Hamiltonians of $A$ and $B$, so
\begin{equation}
    K_A = \order{\frac1\hbar},\qquad K_B = \order{\frac1\hbar}
    \label{Equation: modular Hamiltonian asymptotics}
\end{equation}
for any relevant states $\ket{\psi}\in\mathcal{H}_A\otimes\mathcal{H}_B$. This property will be essential for the holographic interpretation of these states.

But what exactly do we mean by `relevant' states? To answer this, let us first recognise that it is not enough to just assume that the initial state of the combined system is highly entangled. We also need this property to be conserved during the evolution of the system, and this is not immediately guaranteed. One way to ensure it happens is to just assume that the Hamiltonian of the combined system takes the form
\begin{equation}
    H = H_A\otimes I_B + I_A\otimes H_B,
    \label{Equation: decoupled Hamiltonian}
\end{equation}
where $H_A,H_B$ are Hamiltonians for each of the two subsystems, and $I_A,I_B$ are the identities acting on $\mathcal{H}_A,\mathcal{H}_B$ respectively. In other words, there are no interactions between the two subsystems. Under time evolution we would then have
\begin{equation}
    \rho_A(t) = e^{-iH_At/\hbar}\rho_A(0)e^{iH_At/\hbar}, 
    \qquad
    \rho_B(t) = e^{-iH_Bt/\hbar}\rho_B(0)e^{iH_Bt/\hbar},
\end{equation}
which implies
\begin{equation}
    K_A(t) = e^{-iH_At/\hbar}K_A(0)e^{iH_At/\hbar}, 
    \qquad
    K_B(t) = e^{-iH_Bt/\hbar}K_B(0)e^{iH_Bt/\hbar}.
\end{equation}
Since $H_A,H_B$ are assumed to be classical operators, this evolution would preserve the asymptotics of the modular Hamiltonians, so~\eqref{Equation: modular Hamiltonian asymptotics} will hold at all times. 

However, this kind of evolution is far too trivial for our purposes. For example, with this evolution the entanglement entropy
\begin{equation}
    S_A(t) = \tr(\rho_A(t)K_A(t)) = \tr(\rho_A(0)K_A(0))
\end{equation}
would not depend on time. So, comparing with the HRT formula, if an emergent holographic bulk did exist, the area of the HRT surface would be fixed for all time. Similarly, any other quantity which only depended on the density matrix $\rho_A$ up to unitary conjugation would be constant in time. But we would like to allow such quantities, and their purported geometric bulk duals, to fluctuate.

So to get anything interesting we will need some kind of non-trivial dynamical process involving interactions between $A$ and $B$. There are potentially many different processes which have the right properties, but in this paper we will make a particular choice. Let us introduce a third component to our setup: the `environment'. We assume that the environment has a very large number of degrees of freedom and is evolving chaotically. We will not allow $A$ and $B$ to interact directly, but instead couple them both to the environment. Such a coupling can be difficult to analyse in general, but a characteristic phenomenon known as decoherence can occur. We will give a rather brief description of this phenomenon, but the general theory of decoherence is quite subtle. For more information, see for example~\cite{PointerBasis,RobustnessPointerStates,QuantumOriginsofClassical,CollisionalDecoherence,SchlosshauerDecoherence}. Decoherence and chaos in quantum gravity have previously been studied in~\cite{HartleDecoherence,BoundonChaos,ExtremeDecoherence,DecoherenceCFT,DecoherenceSYK}.

Depending on the exact details of the coupling, there is an emergent set of states in $\mathcal{H}_A\otimes\mathcal{H}_B$ known as `pointer' states, and the effect of decoherence is essentially indistinguishable from a projective measurement onto these states. To be more precise, let $\mathcal{M}$ be a space equipped with a measure $\dd{X}$ and a map $X\to\ket{X}$ from $\mathcal{M}$ to $\mathcal{H}_A\otimes\mathcal{H}_B$. The states $\ket{X}$ are the pointer states. We require 
\begin{equation}
    I = \int \dd{X}\ket{X}\bra{X}
\end{equation}
to be a resolution of the identity acting on $\mathcal{H}_A\otimes\mathcal{H}_B$. Suppose the state of $A$ and $B$ is initially given by a joint density matrix $\rho$. Then after decoherence the state becomes
\begin{equation}
    \rho \to \int \dd{X}\mel{X}{\rho}{X}\, \ket{X}\bra{X}.
    \label{Equation: decoherence map}
\end{equation}
This is exactly what happens in a projective measurement onto the pointer states, as we have already stated. The most general possible quantum measurement (described by a positive operator valued measure or POVM) does not require the measurable states to be orthogonal to each other, and the same is true here: in general the pointer states $\ket{X}$ need not be orthogonal to each other. 

We will assume that the evolution of the entire setup is such that $A$ and $B$ periodically come into contact with the environment,\footnote{For intuition, one might picture the environment as a gas of particles, with each particle occasionally colliding with $A$ and $B$.} and that the result of this contact is for the state of $A$ and $B$ to undergo decoherence. Thus, the map~\eqref{Equation: decoherence map} is periodically applied to the state of the system. For simplicity, we will assume that the time period $\Delta t$ between each instance of decoherence is fixed, and moreover that the decoherence itself happens so quickly as to be effectively instantaneous. When decoherence is not happening, we will assume that $A$ and $B$ are not in contact with the environment, and so just evolve unitarily according to some Hamiltonian $H$. So, after a time $\Delta t$, if the state of the system was initially $\rho=\ket{\psi}\bra{\psi}$ it will have evolved into\footnote{We have chosen to write this in such a way that the unitary evolution happens before the decoherence, rather than at any other time. This choice is made without loss of generality. To see this, note that we can redefine the pointer states $\ket{X}\to e^{-iH\Delta t/\hbar}\ket{X}$, and the expression on the right-hand side of~\eqref{Equation: Delta t evolution} will be transformed into one such that decoherence appears to happen before unitary evolution. Similarly, by a redefinition $\ket{X}\to e^{-iHt/\hbar}\ket{X}$ for any $t\in [0,\Delta t]$, we can transform the expression so that decoherence happens at any point during the unitary evolution. Regardless of this redefinition, the time between each occurrence of decoherence in~\eqref{Equation: T evolution} will always be $\Delta t$.}
\begin{equation}
    \rho(\Delta t) = \int\dd{X}\mel{X}{e^{-iH\Delta t/\hbar}}{\psi}\mel{\psi}{e^{iH/\Delta t/\hbar}}{X}\, \ket{X}\bra{X}.
    \label{Equation: Delta t evolution}
\end{equation}
This accounts for both the unitary evolution and the decoherence. To get the evolution of the state after a time $T=n\Delta t$ with $n$ an integer, we can just repeat~\eqref{Equation: Delta t evolution} $n$ times. We end up with
\begin{equation}
    \rho(T) = \int \prod_{l=1}^n \dd{X_l} \abs{\mel{X_1}{e^{-iH\Delta t/\hbar}}{\psi}}^2 \qty(\prod_{k=2}^n\abs{\mel{X_k}{e^{-iH\Delta t/\hbar}}{X_{k-1}}}^2)\, \ket{X_n}\bra{X_n}.
    \label{Equation: T evolution}
\end{equation}
With this, we can compute the transition probability from an initial state $\ket{\psi}$ to a final state $\ket*{\psi'}$:
\begin{equation}
    \mel*{\psi'}{\rho(T)}{\psi'}.
\end{equation}

Actually, in this paper we will be interested in more than just transition probabilities -- we will also consider correlators of operators $O_i$ acting on $\mathcal{H}_A\otimes\mathcal{H}_B$, inserted during the evolution between $\ket{\psi}$ and $\ket{\psi'}$. We assume that we are completely ignorant of the state of the environment, and this is an obstruction to computing these correlators exactly. However, what we can do is compute \emph{expectation values} of correlators by averaging over an appropriate random distribution of environment states. In Appendix~\ref{Appendix: Generating function for decohering system}, we show that these correlator expectation values can be computed with a generating function
\begin{multline}
    Z[\psi,\psi';J] = \int\prod_{l=1}^n\dd{X}_l\braket{\psi'}{X_n}\mel{X_1}{e^{-iJ_1\cdot O_1/\hbar}e^{-iH\Delta t/\hbar}}{\psi}\mel{\psi}{e^{iH\Delta t/\hbar}}{X_1}\braket{X_n}{\psi'}\\
                    \times \prod_{k=2}^n \mel{X_k}{e^{-iJ_k\cdot O_k/\hbar}e^{-iH\Delta t/\hbar}}{X_{k-1}}\mel{X_{k-1}}{e^{iH\Delta t/\hbar}}{X_k}.
                    \label{Equation: decoherence generating function}
\end{multline}
Here $O_k$ are a set of operators inserted at times $t_k=k\Delta t$, and $J_k$ are sources for these operators. The $\cdot$ in $J_k\cdot O_k$ is supposed to denote a sum over all possible operators we want to be able to insert. Note that at $J=0$ the generating function is equal to the transition probability. Also, the correlator expectation value is given by
\begin{equation}
    \expval{O_m(t_m)\dots O_1(t_1)} = \left.\frac{(i\hbar)^m}{Z[\psi,\psi';0]}\pdv{J_m}\dots\pdv{J_1}Z[\psi,\psi';J]\right|_{J=0}.
    \label{Equation: correlator expectation value}
\end{equation}
Despite the coupling between the system and environment, one can observe that the generating function does not depend at all on the fine details of the evolution of the environment. All that is relevant are the set of pointer states $\ket{X}$, the system Hamiltonian $H$, and the sources and operators $J,O$.

\begin{figure}
    \centering
    \begin{tikzpicture}[very thick]
        \everymath\expandafter{\the\everymath\displaystyle}
        \draw[red,dashed] (0,2.5) .. controls (1,1.5) and (7,3.5) .. (8,2.5);
        \draw[red,dashed] (0,5.5) .. controls (1,4.5) and (7,6.5) .. (8,5.5);
        \draw[red,dashed] (0,7) .. controls (1,6) and (7,8) .. (8,7);
        \fill[white] (4,2.5) circle (0.8);
        \fill[white] (4,5.5) circle (0.95);
        \fill[white] (4,7) circle (0.8);
        \node[red] at (4,2.5) {\Large$\int\dd{X_1}$};
        \node[red] at (4,5.5) {\Large$\int\dd{X_{n-1}}$};
        \node[red] at (4,7) {\Large$\int\dd{X_n}$};
        \draw[<->,semithick] (-1.8,5.4) -- (-1.8,7) node[midway,left] {$\Delta t$};
        \draw[dashed,thin] (-1.65,5.5) -- (-1.35,5.5);
        \draw[dashed,thin] (-1.65,6.9) -- (-1.05,6.9);
        \begin{scope}
            \draw (0,1.5) -- (0,3.5);
            \draw[dotted] (0,3.7) -- (0,4.3);
            \draw[->] (0,4.5) -- (0,8);
            \node[rotate=-15,right] at (0,1.5) {$\ket{\psi}$};
            \node[rotate=-15,left] at (0,8) {$\bra{\psi'}$};
            \fill[white] (0,2.5) circle (0.15);
            \fill[white] (0,5.5) circle (0.15);
            \fill[white] (0,7) circle (0.15);
            \fill[red!15] (0,2.5) ellipse (0.15 and 0.25);
            \fill[red!15] (0,5.5) ellipse (0.15 and 0.25);
            \fill[red!15] (0,7) ellipse (0.15 and 0.25);
            \fill (0,1.5) circle (0.08);
            \fill (0,2.35) circle (0.08) node[above right] {$\ket{X_1}$};
            \fill (0,2.65) circle (0.08) node[below left] {$\bra{X_1}$};
            \fill (0,5.35) circle (0.08) node[above right] {$\ket{X_{n-1}}$};
            \fill (0,5.65) circle (0.08) node[below left] {$\bra{X_{n-1}}$};
            \fill (0,6.85) circle (0.08) node[above right] {$\ket{X_n}$};
            \fill (0,7.15) circle (0.08) node[below left] {$\bra{X_n}$};
        \end{scope}
        \begin{scope}[shift={(8,0)}]
            \draw (0,1.5) -- (0,3.5);
            \draw[dotted] (0,3.7) -- (0,4.3);
            \draw[->] (0,4.5) -- (0,8);
            \node[rotate=-15,right] at (0,1.5) {$\ket{\psi}$};
            \node[rotate=-15,left] at (0,8) {$\bra{\psi'}$};
            \fill[white] (0,2.5) circle (0.15);
            \fill[white] (0,5.5) circle (0.15);
            \fill[white] (0,7) circle (0.15);
            \fill[red!15] (0,2.5) ellipse (0.15 and 0.25);
            \fill[red!15] (0,5.5) ellipse (0.15 and 0.25);
            \fill[red!15] (0,7) ellipse (0.15 and 0.25);
            \fill (0,1.5) circle (0.08);
            \fill (0,2.35) circle (0.08) node[above right] {$\ket{X_1}$};
            \fill (0,2.65) circle (0.08) node[below left] {$\bra{X_1}$};
            \fill (0,5.35) circle (0.08) node[above right] {$\ket{X_{n-1}}$};
            \fill (0,5.65) circle (0.08) node[below left] {$\bra{X_{n-1}}$};
            \fill (0,6.85) circle (0.08) node[above right] {$\ket{X_n}$};
            \fill (0,7.15) circle (0.08) node[below left] {$\bra{X_n}$};
        \end{scope}
        \fill[blue] (8,1.9) circle (0.1);
        \fill[blue] (8,4.9) circle (0.1);
        \fill[blue] (8,6.4) circle (0.1);
        \draw[blue] (8,1.9) .. controls (9,1.9) and (9.1,2.1) .. (9.4,2.1)
            node[right] {$J_1\cdot O_1$};
        \draw[blue] (8,4.9) .. controls (9,4.9) and (9.1,5.1) .. (9.4,5.1)
            node[right] {$J_{n-1}\cdot O_{n-1}$};
        \draw[blue] (8,6.4) .. controls (9,6.4) and (9.1,6.6) .. (9.4,6.6)
            node[right] {$J_n\cdot O_n$};
    \end{tikzpicture}
    \caption{The generating function for a system undergoing decoherence involves two copies of the system, one with sources, and the other without. The effect of the decoherence is to project both copies onto the same pointer state $\ket{X_k}$ periodically with time period $\Delta t$. The pointer states are then integrated over.}
    \label{Figure: decoherence generating function contact}
\end{figure}
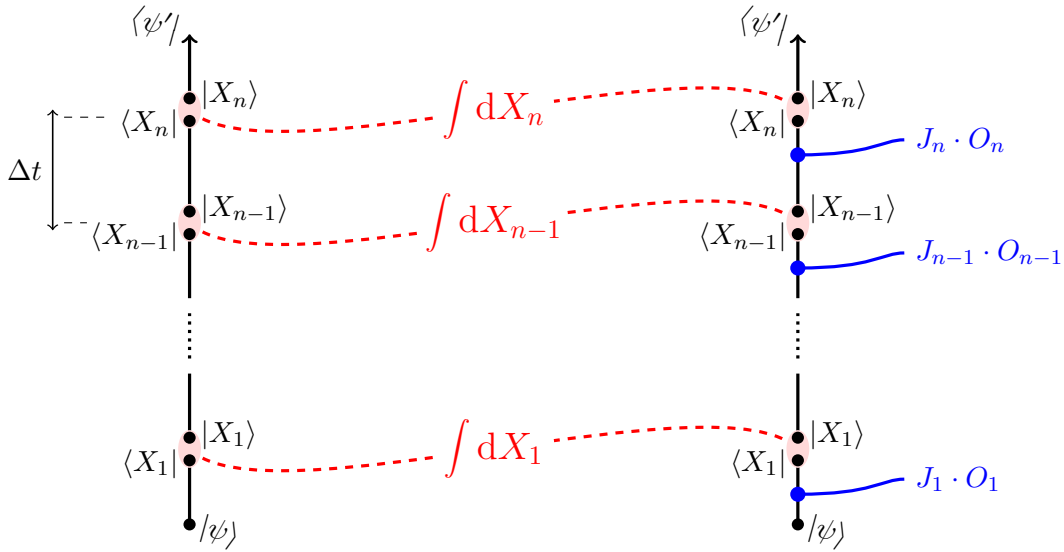

There are two instances of the state undergoing time evolution in this generating function. This is typical of generating functions describing the evolution of open systems, the most commonly encountered example of this being within the Schwinger-Keldysh formalism~\cite{Schwinger,Bakshi1,Bakshi2,Keldysh,SchwingerKeldyshAdSCFT,QuantumDecoherenceHolography,SchwingerKeldyshI,SchwingerKeldyshII,SchwingerKeldyshHolograms}. The effect of the environment usually manifests in the generating function as a kind of interaction between the two instances, and a common way to handle this is in terms of a Feynman-Vernon `influence functional'~\cite{FeynmanVernon}. Here the interaction is expressed slightly differently: the two instances evolve more or less independently, except for when decoherence happens. The effect of the decoherence is to effectively bring the two instances into contact, by projecting both onto the \emph{same} pointer state. This is depicted in Figure~\ref{Figure: decoherence generating function contact}, which also portrays the fact that operators are only inserted in one of the instances. 

From now on we will just refer to~\eqref{Equation: correlator expectation value} as a correlator, but it is worth keeping in mind that it is really an expectation value of a correlator. It has recently been argued that such expectation values of correlators are an essential feature of \emph{any} kind of holographic theory~\cite{TranscendingTheEnsemble}.

For a system experiencing decoherence, the pointer states $\ket{X}$ should be viewed as the macroscopic classical states of the system. In the lab, decoherence is essentially the reason why it is difficult to set up long-lived superpositions of such states -- the coupling with the environment quickly destroys the superposition, converting it to correlations with the environment that rapidly disperse. In our case, we should think of the pointer states as the classical states of the \emph{bulk} theory. Thus, the pointer states are the `relevant' states described previously, so we will assume that they are highly entangled in the classical limit, in the sense given above.

We will make a further assumption about the structure of the pointer states, inspired again by holography. Suppose $U_A\in U(\mathcal{H}_A)$ and $U_B\in U(\mathcal{H}_B)$ are unitary operators acting on $\mathcal{H}_A$ and $\mathcal{H}_B$ respectively. Then if $\ket{X}$ is one pointer state, we assume for all such $U_A,U_B$ that $\ket{X'}$ is also a pointer state, where
\begin{equation}
    \ket{X'} = \big(U_A\otimes U_B\big)\ket{X}.
\end{equation}
Moreover, we assume that the pointer state measure $\dd{X}$ is invariant under this action of $U(\mathcal{H}_A)\times U(\mathcal{H}_B)$. One way of interpreting this is as follows. $U_A\otimes U_B$ is an operator which changes only the degrees of freedom which are local to $A$ and $B$. Bulk holographic states should share all the local symmetries of the boundary states. In other words, if we take a bulk state, and apply a local operator at the boundary, we should get back another bulk state, which is why we need $\ket{X'}$ to be a pointer state. The invariance of $\dd{X}$ means that decoherence does not affect the local degrees of freedom. Instead, it only has an impact on non-local degrees of freedom encoded in the structure of the entanglement between the two systems.

Let us summarise the assumptions we have made so far.
\begin{itemize}
    \item \textbf{Decoherence happens, and it happens quickly and frequently.} We assume that the quantum theory under consideration consists of two components: a `system' and an `environment'. We assume that the environment evolves in such a way that the coupling between the two components leads to the system experiencing decoherence onto a set of pointer states $\ket{X}$. We also assume that the decoherence happens so quickly as to be effectively instantaneous, and that it happens frequently, with a time period $\Delta t$ between each occurrence. The simple model described above will be the one we use to analyse the resulting evolution.
    \item \textbf{The system is made of two subsystems with classical limits.} We assume that the system Hilbert space decomposes as $\mathcal{H}=\mathcal{H}_A\otimes\mathcal{H}_B$, where $\mathcal{H}_A, \mathcal{H}_B$ are the Hilbert spaces of two subsystems $A$ and $B$. Furthermore, we require these individual subsystems to have a classical limit $\hbar\to 0$ in terms of coherent states.
    \item \textbf{The decohering process is compatible with the local symmetries of $A$ and $B$.} By this we mean the pointer states $\ket{X}$ and measure $\dd{X}$ are invariant under the action of $U(\mathcal{H}_A)\times U(\mathcal{H}_B)$, in the way just described.
    \item \textbf{The pointer states are highly entangled in the classical limit.} Let
        \begin{equation}
            \rho_A(X)=\tr_B\ket{X}\bra{X}, \qquad
            \rho_B(X)=\tr_A\ket{X}\bra{X}
            \label{Equation: reduced pointer states}
        \end{equation}
        be the reduced density matrices of the pointer state $\ket{X}$ in subsystems $A,B$ respectively. We will assume that $\rho_A$ and $\rho_B$ are invertible, and that their corresponding modular Hamiltonians
        \begin{equation}
            K_A(X) = -\log\rho_A(X), \qquad
            K_B(X) = -\log\rho_B(X)
        \end{equation}
        obey
        \begin{equation}
            K_A = \order{\frac1{\hbar}}, \qquad
            K_B = \order{\frac1{\hbar}}.
        \end{equation}
\end{itemize}

In addition to the classical limit $\hbar\to 0$, we will take another limit in this paper: $\Delta t \to 0$. This essentially means that decoherence happens very frequently. It is important to carefully specify the relative scaling of $\hbar,\Delta t$, so that the simultaneous limit $\Delta t,\hbar \to 0$ has a well-defined outcome. Different phenomena will arise in different scaling regimes, but for us the most interesting physics will happen when the scaling obeys our final key assumption:
\begin{itemize}
    \item \textbf{$\hbar$ and $\Delta t$ obey a certain scaling relationship.} In the limit $\Delta t,\hbar \to 0$, we require
        \begin{equation}
            \frac{\Delta t}{\hbar}\to 0, \qquad \frac{\Delta t}{\hbar^2} \to \infty.
            \label{Equation: Delta t hbar scaling}
        \end{equation}
        Thus, decoherence happens very frequently in the classical limit, but not too frequently.
\end{itemize}
It is worth noting the particular meaning of `small' for each of these quantities: small $\hbar$ means $\hbar$ is much smaller than the classical action, and small $\Delta t$ means $\Delta t \ll T$ where $T$ is the timescale on which we are making observations. 

There are a few other assumptions we will make in this paper. However, we view these as less essential, and our main reason for making them is to avoid overcomplicating the calculations. We expect (or hope) that they could be dropped, and the mechanism we describe in this paper would still work in a generalised form.\footnote{Of course, it may also be true that the previous `essential' assumptions can be weakened.} First, we will consider the case where the two subsystems $A$ and $B$ are actually just two copies of the same system, so that $\mathcal{H}_A=\mathcal{H}_B$ and the coherent states are the same. For notational purposes, we will in some cases continue to use subscripts ${}_{A,B}$ to label the subsystems. Second, the Hamiltonian $H$ can in general be written as
\begin{equation}
    H = H_A\otimes I_B + I_A\otimes H_B + H_{\text{int}},
    \label{Equation: system Hamiltonian}
\end{equation}
where $H_{\text{int}}$ is an interaction term. We will assume that $H_{\text{int}}=0$, so that when decoherence is not happening there are no interactions between the two systems. It would be interesting in the future to allow $H_{\text{int}}\ne 0$, to see if one could reproduce the results of~\cite{TraversableWormholes}. Third, we will consider only correlators of operators acting on one of the subsystems -- without loss of generality let that subsystem be $A$. This means the operators can be written $O\otimes I_B$. A final convenient assumption can be made without loss of generality: we choose the sources and operators such that $J\cdot O$ is Hermitian. This can always be made to be true by an appropriate linear redefinition of the sources.

We will argue in this paper that the assumptions described above have the following consequence. Suppose the semiclassical physics of $A$ and $B$ when they are isolated from the environment is $d$-dimensional. By this we mean that all correlators of quantum operators around a classical background may be computed in terms of a field theory living in $d$ dimensions. Then the semiclassical physics of the systems coupled to the environment is $(d+1)$-dimensional. Moreover, the fields have non-trivial dynamics in the extra dimension, so this gives a genuine example of emergent holography. One attractive feature of the additional dimension is that it is generated by modular flow, in line with previous research~\cite{BulkLocalityModularFlow}. 

We should emphasise that we have not tried to find an example of a fundamental theory that satisfies all the assumptions. However, given the genericity of the assumptions, we would be very surprised if such a theory did not exist. Additionally, a major point in favour of the assumptions is the resulting mechanism that we describe in this paper. Although several of the assumptions were inspired by holography, it is far from obvious that they were enough to lead to \emph{actual} holograhy. In this sense, the consequences we describe are greater than the sum of the assumptions, and so are worth studying, a posteriori.

Let us provide a roadmap for the rest of the paper.

First, in Section~\ref{Section: Polar decomposition of pointer states} we will describe a useful parametrisation of the pointer states $\ket{X}$ in terms of the reduced state $\rho_A=\tr\ket{X}\bra{X}$, and a unitary operator $U$. Our assumptions will imply that $\rho_A$ and $U$ can be independently specified, and this allows us to consider their dynamics separately.

Next, in Section~\ref{Section: Frequent decoherence}, we will consider the limit in which decoherence happens very frequently, i.e.\ $\Delta t \to 0$. This will mean that there will be a very large number of pointer states that we integrate over in the generating function, and we end up with an integral over paths of pointer states. We will obtain an expression for the `action' of this path integral in terms of the variables $\rho_A$ and $U$, and show that it involves a dynamical generalisation of Uhlmann holonomy~\cite{Uhlmann:sp,Uhlmann1991,Uhlmann1992}, which is a notion of parallel transport of $U$ along the path of density matrices $\rho_A$, and the Bures metric~\cite{Braunstein,Jozsa:1994,Bures,Helstrom}, which is a metric on the space of density matrices $\rho_A$. We will describe these further in Sections~\ref{Section: Dynamical Uhlmann holonomy} and~\ref{Section: Bures metric}.

We will then consider the classical limit $\hbar\to 0$ in Section~\ref{Section: Classical limit}, showing that the terms involving the dynamical Uhlmann holonomy scale like $1/\hbar^2$, while the terms involving the Bures metric scale like $1/\hbar$. We will use this to show that, in the combined limit $\Delta t,\hbar\to 0$, the path of density matrices $\rho_A$ must fall within a certain class of paths which includes all differentiable paths, while the path of unitary operators $U$ must follow the dynamical Uhlmann holonomy along $\rho_A$. This provides an alternate perspective on the results of~\cite{Kirklin4}.

The reduced state $\rho_A$ has no knowledge of the dynamical Uhlmann holonomy, so one might wonder whether the Uhlmann holonomy is actually observable from the point of view of system $A$. After all, expectation values
\begin{equation}
    \expval{O} = \tr(\rho_A O)
\end{equation}
of operators $O$ acting on $A$ clearly do not depend on the holonomy. In fact, the holonomy is observable. There is an indirect coupling between systems $A$ and $B$ through the environment. This coupling means that evolution of the system in $B$ can affect the evolution of the system in $A$, and vice versa. Thus, the correlator of two operators $O_1,O_2$ acting on $A$ at two different times $t_1< t_2$
\begin{equation}
    \expval{O_2(t_2)O_1(t_1)}
\end{equation}
depends on the state in $B$. In particular, the insertion of $O_1$ at $t_1$ will result in changes that propagate into $B$, and then back into $A$, where they will be detected by $O_2$ at $t_2$. Thus, these correlators are sensitive to the Uhlmann holonomy. In Section~\ref{Section: Semiclassical correlators} we will obtain a path integral formula for correlators of an arbitrary number of operators using a generating function. In this way we will explicitly show how the correlators depend on the dynamical Uhlmann holonomy.

Section~\ref{Section: Emergent holography} is the crux of the paper. In it we will demonstrate that the generating function obtained in Section~\ref{Section: Semiclassical correlators} is secretly a holographic one. We will do this by deriving a path integral formula for the dynamical Uhlmann holonomy in terms of coherent states. The action for this path integral involves one more dimension than the original action~\eqref{Equation: isolated action} of the coherent states.

We conclude the paper in Section~\ref{Section: Discussion} with some speculation on future directions.

\section{Highly entangled decohering systems}
\label{Section: Highly entangled decohering systems}

\subsection{Polar decomposition of pointer states}
\label{Section: Polar decomposition of pointer states}

Consider a pointer state $\ket{X} \in \mathcal{H}_A\otimes\mathcal{H}_B$. By dualising on the $\mathcal{H}_B$ part, we can view this state as a linear map $W_X:\mathcal{H}_B\to\mathcal{H}_A$. The reduced state in $A$ is given by
\begin{equation}
    \rho_A(X) = \tr_B \ket{X}\bra{X} = W_X W_X^\dagger.
\end{equation}
Let $\mathcal{M}_A$ denote the space of all such reduced states for all $X\in\mathcal{M}$. For notational simplicity we will drop the subscript ${}_A$ on $\rho_A$, i.e.\ just write $\rho =\rho_A$. Because we are assuming $\mathcal{H}_A=\mathcal{H}_B$, $W_X$ is really a map from $\mathcal{H}_A$ to itself, so we can do a polar decomposition of $W_X$ to get
\begin{equation}
    W_X = \sqrt{W_X W_X^\dagger} U^\dagger = \sqrt{\rho(X)}U^\dagger,
    \label{Equation: pointer state polar decomposition}
\end{equation}
where $U\in U(\mathcal{H}_A)$ is some unitary operator. Here the square root $\sqrt{\rho}$ is the unique positive operator satisfying $(\sqrt{\rho})^2=\rho$; this exists because $\rho$ is positive. We are assuming $\rho(X)$ is invertible, so this polar decomposition is unique, i.e.\ $U$ is uniquely determined. Moreover, because we are assuming compatibility of the pointer states with the local symmetries of $A$ and $B$, 
\begin{equation}
    W_{X'} = W_X {U'}^\dagger = \sqrt{\rho(X)}U^\dagger {U'}^\dagger = \sqrt{\rho(X)} (U'U)^\dagger
    \label{Equation: pointer state unitary action}
\end{equation}
is also a pointer state, for any unitary operator $U'\in U(\mathcal{H}_A)$. This gives the space of pointer states $\mathcal{M}$ the structure of a $U(\mathcal{H}_A)$-principal bundle, where the base space is the space $\mathcal{M}_A$ of all reduced pointer states $\rho(X)$, the projection map is $X\to \rho(X)$, and the fibre over $\rho(X)$ is given by all states of the form~\eqref{Equation: pointer state polar decomposition}. Because $\dd{X}$ is invariant under the action \eqref{Equation: pointer state unitary action} of $U(\mathcal{H}_A)$, we can decompose it as 
\begin{equation}
    \dd{X} = \dd{\rho}\,\dd{U},
\end{equation}
where $\dd{\rho}$ is a measure of integration over the space of fibres (with each fibre labelled by the reduced pointer state $\rho$), and $\dd{U}$ is the invariant measure on $U(\mathcal{H}_A)$.

In the polar decomposition~\eqref{Equation: pointer state polar decomposition}, the part of the state in subsystem $A$ is completely accounted for by the factor involving the density matrix $\rho(X)$. By this, we mean that all expectation values of operators acting on $A$ only depend on $\rho(X)$, and not $U$. Clearly, therefore, $U$ must account for everything else, including the state in $B$, as well as some details of the entanglement between $A$ and $B$. 

Consider the generating function~\eqref{Equation: decoherence generating function} for evolution from an initial pointer state $\ket{X_0}$ to a final pointer state $\ket{X_{n+1}}$ in the presence of sources $J$ after a time $T=n\Delta t$. This may be written
\begin{equation}
    Z[X_0,X_{n+1};J] = \int\prod_{k=1}^n\qty(\dd{X}_k \mel{X_k}{e^{-iJ_k\cdot O_k/\hbar}e^{-iH\Delta t/\hbar}}{X_{k-1}}\mel{X_{k-1}}{e^{iH\Delta t/\hbar}}{X_k})\abs{\braket{X_{n+1}}{X_n}}^2 .
\end{equation}
We are only considering operators which act on $\mathcal{H}_A$, so we can replace $O_k\to O_k\otimes I_B$. Also, whenever we compute correlators or transition probabilities we eventually set $J=0$, so we can assume $J$ is arbitrarily small, and use this to rescale $J\to \Delta t J$. Finally, we are assuming that $H_{\text{int}}=0$ in~\eqref{Equation: system Hamiltonian}. Thus, using the polar decomposition~\eqref{Equation: pointer state polar decomposition}, we can write the factors in the generating function as
\begin{align}
    \mel{X_k}{e^{-iJ_k\cdot O_k\otimes I_B\Delta t/\hbar}e^{-iH\Delta t/\hbar}}{X_{k-1}}
    &=
    \tr(U_k\sqrt{\rho_k}e^{-iJ_k\cdot O_k\Delta t/\hbar}e^{-iH_A\Delta t/\hbar}\sqrt{\rho_{k-1}}U_{k-1}^\dagger e^{iH_B\Delta t/\hbar}),\\
    \mel{X_{k-1}}{e^{iH\Delta t/\hbar}}{X_k} 
    &=
    \tr(U_{k-1}\sqrt{\rho_{k-1}}e^{iH_A\Delta t/\hbar}\sqrt{\rho_k}U_k^\dagger e^{-iH_B\Delta t/\hbar}).
\end{align}
where we have written the pointer states $\ket{X_k}$ as linear maps $W_{X_k} = \sqrt{\rho(X_k)}U_k^\dagger$, and set $\rho_k=\rho(X_k)$. In terms of these variables, we can therefore write
\begin{equation}
    Z[X_0,X_{n+1};J] = \int\prod_{k=1}^n\Big(\dd{\rho_k}\dd{U_k} Y_k\Big)\abs{\tr(U_{n+1}\sqrt{\rho_{n+1}}\sqrt{\rho_n}U^\dagger_n)}^2.
    \label{Equation: generating function polar}
\end{equation}
where
\begin{multline}
    Y_k = \tr(U_k\sqrt{\rho_k}e^{-iJ_k\cdot O_k\Delta t/\hbar}e^{-iH_A\Delta t/\hbar}\sqrt{\rho_{k-1}}U_{k-1}^\dagger e^{iH_B\Delta t/\hbar})\\
    \times
    \tr(U_{k-1}\sqrt{\rho_{k-1}}e^{iH_A\Delta t/\hbar}\sqrt{\rho_k}U_k^\dagger e^{-iH_B\Delta t/\hbar})
\end{multline}

\subsection{Frequent decoherence\texorpdfstring{ ($\Delta t\to 0$)}{}}
\label{Section: Frequent decoherence}

Consider the limit $\Delta t \to 0$, keeping $T$ approximately fixed, so that the integer $n$ becomes very large. Then the generating function~\eqref{Equation: generating function polar} takes on the characteristics of a path integral. In particular, it is dominated by those sequences $\rho_k,U_k$ for which each of the factors in the integrand is near to its maximum. Such sequences can be approximated as points along continuous paths $\rho_k = \rho(t_k)$ and $U_k=U(t_k)$, with $t_k=k\Delta t$.

The typical paths which contribute to the path integral are not differentiable, but instead obey
\begin{equation}
    \ket{X_k}-\ket{X_{k-1}} = \order{\sqrt{\Delta t}}
    \label{Equation: Wiener}
\end{equation}
with measure 1 (with regard to the path integral measure). The space of paths with this behaviour is sometimes called an abstract Wiener space, and the structure theorem for Gaussian measures essentially says that all path integrals must be done over such a space. Given~\eqref{Equation: Wiener}, it is shown in Appendix~\ref{Appendix: Y_k in exponential form} that $Y_k$ then takes the form
\begin{multline}
    Y_k = \exp\Big[\tr(-\delta\qty(\sqrt{\rho_k})\delta\qty(\sqrt{\rho_k})-\qty(\sqrt{\rho_k}\delta\qty(\sqrt{\rho_k})-\delta\qty(\sqrt{\rho_k})\sqrt{\rho_k})C_k + \tilde\rho_k C_k^2) - \tr(\tilde\rho_k C_k)^2 \\
    +\frac{i\Delta t}{\hbar}\tr(\qty(H_A+\frac12J_k\cdot O_k)\big(\delta(\sqrt{\rho_k})\sqrt{\rho_k}-\sqrt{\rho_k}\delta(\sqrt{\rho_k})-2\sqrt{\rho_k}C_k\sqrt{\rho_k}\big))\\
    + \frac{2i\Delta t}{\hbar}\tr(\rho_k\qty(H_A+\frac12 J_k\cdot O_k))\tr(\tilde\rho_kC_k)
    - \frac{i\Delta t}{\hbar} \tr(\tilde\rho_k J_k\cdot O_k) +\order{\Delta t^2}\Big],
    \label{Equation: Y_k exponential}
\end{multline}
where $\tilde\rho_k = \frac12\qty(\rho_k+\rho_{k-1})$ and
\begin{equation}
    C_k = \frac12\qty(U_{k-1}^\dagger e^{iH_B\Delta t/\hbar}U_k-U_k^\dagger e^{-iH_B\Delta t/\hbar}U_{k-1}) = \order{\sqrt{\Delta t}}.
    \label{Equation: C_k scaling}
\end{equation}
This can be inverted to get
\begin{equation}
    U_{k-1}^\dagger e^{iH_B\Delta t/\hbar} U_k = \exp(C_k+\order{\Delta t^{3/2}}).
    \label{Equation: C_k inverse}
\end{equation}
In \eqref{Equation: Y_k exponential} and the following, the symbol $\delta$ is defined such that $\delta q_k = q_k-q_{k-1}$ for any quantity $q_k$ with an index $k\in1,\dots{n+1}$.

%

The exponent in~\eqref{Equation: Y_k exponential} is essentially just a complicated quadratic in $C_k$, and we will now complete the square. To find the stationary point, we can consider a linearised variation $C_k\to C_k+\Delta C_k$. Under such a variation, the exponent changes by
\begin{multline}
    \tr(\Delta C\qty[\tilde{\rho_k}C_k+C_k\tilde{\rho_k} - \sqrt{\rho_k}\delta(\sqrt{\rho_k})+\delta(\sqrt{\rho_k})\sqrt{\rho_k}-\frac{2i\Delta t}{\hbar}\sqrt{\rho_k}\qty(H_A+\frac12J_k\cdot O_k)\sqrt{\rho_k}])\\
    - 2\tr(\tilde\rho_k \Delta C_k)\tr(\tilde\rho_kC_k - \frac{i\Delta t}{\hbar}\rho_k\qty(H_A+\frac12 J_k\cdot O_k)) + \order{\Delta t^2}. 
\end{multline}
At the maximum, this must vanish for arbitrary $\Delta C_k$, so we must have
\begin{multline}
    \tilde{\rho_k}C_k+C_k\tilde{\rho_k} - \sqrt{\rho_k}\delta(\sqrt{\rho_k})+\delta(\sqrt{\rho_k})\sqrt{\rho_k}-\frac{2i\Delta t}{\hbar}\sqrt{\rho_k}\qty(H_A+\frac12J_k\cdot O_k)\sqrt{\rho_k}\\
    = 2 \tilde\rho_k\tr(\tilde\rho_kC_k - \frac{i\Delta t}{\hbar}\rho_k\qty(H_A+\frac12 J_k\cdot O_k)) + \order{\Delta t^{3/2}}. 
    \label{Equation: C eom}
\end{multline}
The final term is $\order{\Delta t^{3/2}}$ because we have `factored out' $\Delta C=\order{\sqrt{\Delta t}}$. Later we will solve~\eqref{Equation: C eom}, but for now suffice it to say that its solutions are of the form
\begin{equation}
    C_k = a_k + i \sigma_k + \order{\Delta t^{3/2}},
\end{equation}
where $a_k=\order{\sqrt{\Delta t}}$ is the unique fixed anti-Hermitian operator which solves the simpler equation
\begin{equation}
    \tilde{\rho}_ka_k+a_k\tilde{\rho}_k - \sqrt{\rho_k}\delta(\sqrt{\rho_k})+\delta(\sqrt{\rho_k})\sqrt{\rho_k}-\frac{2i\Delta t}{\hbar}\sqrt{\rho_k}\qty(H_A+\frac12J_k\cdot O_k)\sqrt{\rho_k} = 0,
    \label{Equation: a_k equation}
\end{equation}
and $\sigma_k$ is any real number. The freedom in $\sigma_k$ comes from the fact that $Y_k$ is invariant under $U_k\to e^{i\sigma_k}U_k$, which is just a reflection of the usual phase ambiguity in the physical state of a quantum system.

Armed with this solution, we can now actually complete the square, obtaining
\begin{multline}
    Y_k = \exp\Big[\tr(-\delta(\sqrt{\rho_k})\delta(\sqrt{\rho_k}) - \frac12a_k\big(\sqrt{\rho_k}\delta(\sqrt{\rho_k})-\delta(\sqrt{\rho_k})\sqrt{\rho_k}\big))\\
    + \tr(\rho_k(C_k-a_k)^2) - \tr(\rho_k(C_k-a_k))^2 - \frac{i\Delta t}{\hbar}\tr(\rho_k J_k\cdot O_k) + \order{\Delta t^{3/2}}\Big].
    \label{Equation: Y_k completed square}
\end{multline}
It suffices at this point to compute $Y_k$ to this order. 

The final term in the generating function can be written similarly by just setting $J=H_A=H_B=0$. One obtains
\begin{multline}
    \abs{\tr(U_{n+1}\sqrt{\rho_{n+1}}\sqrt{\rho_n}U^\dagger_n)}^2 = \\
    \exp\Big[\tr\Big(-\delta(\sqrt{\rho_{n+1}})\delta(\sqrt{\rho_{n+1}}) - \frac12a_{n+1}\big(\sqrt{\rho_{n+1}}\delta(\sqrt{\rho_{n+1}})-\delta(\sqrt{\rho_{n+1}})\sqrt{\rho_{n+1}}\big)\Big) \\
    + \tr(\rho_{n+1}(C_{n+1}-a_{n+1})^2)
    - \tr(\rho_{n+1}(C_{n+1}-a_{n+1}))^2 + \order{\Delta t^{3/2}}\Big],
    \label{Equation: final term completed square}
\end{multline}
where $C_{n+1} = \frac12(U^\dagger_nU_{n+1}-U^\dagger_{n+1}U_n)$, and $a_{n+1}$ satisfies
\begin{equation}
    \tilde{\rho}_{n+1}a_{n+1}+a_{n+1}\tilde{\rho}_{n+1} - \sqrt{\rho_{n+1}}\delta(\sqrt{\rho_{n+1}})+\delta(\sqrt{\rho_{n+1}})\sqrt{\rho_{n+1}} = 0.
\end{equation}

Substituting~\eqref{Equation: Y_k completed square} and~\eqref{Equation: final term completed square} into the generating function~\eqref{Equation: generating function polar} gives
\begin{equation}
    Z[X_0,X_n;J] = \int\prod_{k=1}^n\big(\dd{\rho_k}\dd{U_k}\big) \exp(-\mathcal{S}),
    \label{Equation: generating function 1}
\end{equation}
where the `action' is
\begin{equation}
    \mathcal{S} = \sum_{k=1}^{n+1}\qty(D_k - \big(\Delta(C_k-a_k)\big)^2 + \frac{i\Delta t}{\hbar}\tr(\rho_k J_k\cdot O_k)) + \order{\sqrt{\Delta t}}.
    \label{Equation: generating action}
\end{equation}
Here
\begin{equation}
    D_k = \tr(\delta(\sqrt{\rho_k})\delta(\sqrt{\rho_k}) + \frac12 a_k\qty(\sqrt{\rho_k}\delta(\sqrt{\rho_k})-\delta(\sqrt{\rho_k})\sqrt{\rho_k})),
\end{equation}
and
\begin{equation}
    \big(\Delta(C_k-a_k)\big)^2 = \tr(\rho_k(C_k-a_k)^2) - \tr(\rho_k(C_k-a_k))^2,
\end{equation}
and we have set $J_{n+1}=0$ to make the notation convenient. $\big(\Delta(C_k-a_k)\big)^2$ is the variance of $C_k-a_k$ in the state $\rho_k$.

In the limit $\Delta t\to 0$, we may discard the $\order{\sqrt{\Delta t}}$ part of~\eqref{Equation: generating action}. 
We end up with
\begin{equation}
    \mathcal{S} = \lim_{\Delta t \to 0}\sum_{k=1}^{n+1}\qty(D_k - \big(\Delta(C_k-a_k)\big)^2 + \frac{i\Delta t}{\hbar}\tr(\rho_k J_k\cdot O_k)).
    \label{Equation: generating action 2}
\end{equation}

\subsection{Dynamical Uhlmann holonomy}
\label{Section: Dynamical Uhlmann holonomy}

\begin{figure}
    \centering
    \begin{tikzpicture}
        \draw[semithick,fill=black!5,rounded corners] (0,0) -- (0,8) -- (3,5.6) -- (3,2) -- cycle;
        \draw[line width=2pt, blue] (0.6,3.85) coordinate (a0)
            arc (180:90:0.9 and 2) coordinate (a1)
            arc (90:10:0.9 and 2) coordinate (a2)
            arc (10:0:0.9 and 2)
            arc (0:-90:0.55 and 1.7) coordinate (a3)
            arc (-90:-140:0.55 and 1.7) coordinate (rho)
            arc (-140:-180:0.55 and 1.7)
            arc (-180:-330:0.3 and 1.3) coordinate (a4);

        \path (a4) --+ (0.1,0.2) coordinate (a4a);
        \path (a4) --+ (-0.15,0.15) coordinate (a4b);
        \draw[line width=2pt, blue] (a4a) -- (a4) -- (a4b);

        \path[line width=2pt, red] (4.6,3.85) coordinate (b0)
            arc (180:90:2.1 and 2) coordinate (b1)
            arc (90:25:2.3 and 2.8) coordinate (b2)
            arc (25:0:2.3 and 2.8) 
            arc (0:-90:0.2 and 0.9) coordinate (b3)
            arc (-90:-155:0.2 and 0.9) coordinate (psi)
            arc (-155:-180:0.2 and 0.9) 
            arc (-180:-270:1.15 and 2) 
            arc (-270:-325:0.7 and 1.3) coordinate (b4);

        \draw[semithick,dashed] (a0) -- (b0);
        \draw[semithick,dashed] (a1) -- (b1);
        \draw[semithick,dashed] (a3) -- (b3);

        \path (4.6,3.85) 
            arc (180:90:2.1 and 2) coordinate (c);
        \draw[line width=2pt, red] (c)
            arc (90:0:2.3 and 2.8) 
            arc (0:-155:0.2 and 0.9);
        \draw[semithick,dashed] (a4) -- (b4);
        \draw[line width=5pt, white] (4.6,3.85) 
            arc (180:90:2.1 and 2);
        \draw[line width=2pt, red] (4.6,3.85) 
            arc (180:90:2.1 and 2);
        \draw[line width=5pt, white] (psi)
            arc (-155:-180:0.2 and 0.9)
            arc (-180:-270:1.15 and 2) 
            arc (-270:-325:0.7 and 1.3);
        \draw[line width=2pt, red] (psi)
            arc (-155:-180:0.2 and 0.9)
            arc (-180:-270:1.15 and 2) 
            arc (-270:-325:0.7 and 1.3) coordinate (b5);

        \path (b5) --+ (0.1,0.2) coordinate (b5a);
        \path (b5) --+ (-0.2,0.1) coordinate (b5b);
        \draw[line width=2pt, red] (b5a) -- (b5) -- (b5b);

        \fill[blue] (rho) circle (0.1);
        \fill[red] (psi) circle (0.1);

        \node[below left] at (rho) {\Large$\rho$};
        \node[left] at (psi) {\Large$\ket{\psi}$};

        \node[left] at (-0.1,3) {\parbox{3cm}{\centering space of reduced states \\ (base space)}};
        \draw[very thick,->] (-1.6,3.9) .. controls (-1.5,5) and (-0.6,5.7) .. (-0.1,5.4);

        \begin{scope}[shift={(0,0.8)}]
            \node[below] at (7,-0.1) {\parbox{5cm}{\centering space of purifications \\ (total space)}};
            \draw[rounded corners, very thick] (7,0) -- (7,0.2) -- (3,0.2) -- (3,0.4);
            \draw[rounded corners, very thick] (7,0) -- (7,0.2) -- (11,0.2) -- (11,0.4);
        \end{scope}
    \end{tikzpicture}
    \caption{The space of density matrices acting on a Hilbert space $\mathcal{H}$ may be viewed as the base space of a fibre bundle whose fibre over $\rho$ consists of all purifications of $\rho$. Uhlmann holonomy provides us with a notion of parallel transport in this bundle, i.e.\ it gives us a way to `lift' a curve of density matrices in the base space to a curve in the full bundle of purifications.}
    \label{Figure: Uhlmann holonomy}
\end{figure}
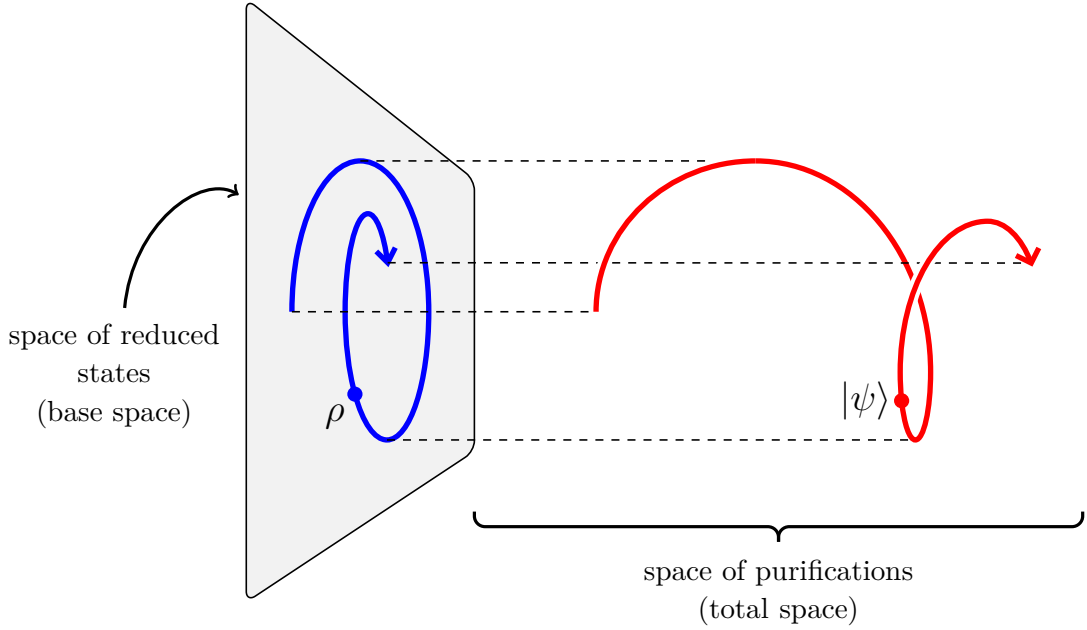

In the previous section, we defined a sequence of operators $a_k$ associated with a sequence of density matrices $\rho_k$. These operators might seem obscure, but actually they are a natural generalisation of an idea due to Uhlmann which has been given the name `Uhlmann holonomy'~\cite{Uhlmann:sp,Uhlmann1991,Uhlmann1992}.
One considers a curve $\rho(t)$ of density matrices acting on a Hilbert space $\mathcal{H}$, and an initial purification $\ket{\psi(0)}$ of $\rho(0)$, i.e.\ a state in an extended Hilbert space $\mathcal{H}\otimes\mathcal{H}'$ obeying
\begin{equation}
    \rho(0) = \tr'\ket{\psi(0)}\bra{\psi(0)}.
\end{equation}
If $\dim(\mathcal{H})\le \dim(\mathcal{H}')$, then there are many ways to extend $\ket{\psi(0)}$ to a curve $\ket{\psi(t)}$ of states in $\mathcal{H}\otimes\mathcal{H}'$, such that $\ket{\psi(t)}$ is a purification of $\rho(t)$ for all $t$. Let us assume that $\mathcal{H}=\mathcal{H}'$, and the density matrices $\rho(t)$ are invertible. Uhlmann holonomy then defines a \emph{unique} such curve $\ket{\psi(t)}$ (up to phase shifts) with a certain privileged property described below. In this way it provides a notion of parallel transport of the purification $\ket{\psi}$ around the curve of density matrices. This is shown in Figure~\ref{Figure: Uhlmann holonomy}.

The property defining Uhlmann holonomy is as follows. Suppose we pick some sequence $\rho_k=\rho(t_k)$, $k=0,\dots, n-1$ of density matrices ordered along the curve, and let $\ket{\psi_k}=\ket{\psi(t_k)}$ be the associated purifications. Then, under Uhlmann holonomy, the curve $\ket{\psi(t)}$ must maximise the `transition probability'
\begin{equation}
    \abs{\braket{\psi_{n+1}}{\psi_n}}^2 \dots \abs{\braket{\psi_2}{\psi_1}}^2 \abs{\braket{\psi_1}{\psi_0}}^2
    \label{Equation: Uhlmann transition probability}
\end{equation}
in the limit as $n\to\infty$ and the density matrices $\rho_k$ densely cover $\rho(t)$.

Let us write the states $\ket{\psi_k}$ as linear maps $\mathcal{H}'\to\mathcal{H}$ by dualising on the $\mathcal{H}'$ part. In fact, let us write them in terms of a polar decomposition, so that $\ket{\psi_k}\sim \sqrt{\rho_k}U_k^\dagger$ for some unitary operator $U_k$. Then~\eqref{Equation: Uhlmann transition probability} takes the form
\begin{equation}
    \abs{\tr(U_{n+1}\sqrt{\rho_{n+1}}\sqrt{\rho_n}U^\dagger_n)}^2\dots\abs{\tr(U_1\sqrt{\rho_1}\sqrt{\rho_0} U_0^\dagger)}^2.
    \label{Equation: Uhlmann transition probability 2}
\end{equation}
One should recognise this as exactly the integrand in the generating function~\eqref{Equation: generating function polar}, but with $J=H_A=H_B=0$.

Another polar decomposition makes it easy to maximise~\eqref{Equation: Uhlmann transition probability 2}. In particular, let us write
\begin{equation}
    \sqrt{\rho_{k+1}}\sqrt{\rho_k} = V_{k+1} \sqrt{\sqrt{\rho_k}\rho_{k+1}\sqrt{\rho_k}},
\end{equation}
where $V_{k+1}$ is some unitary operator. Since $\rho_k,\rho_{k+1}$ are invertible, $V_{k+1}$ is uniquely determined. Since $\sqrt{\sqrt{\rho_k}\rho_{k+1}\sqrt{\rho_k}}$ is a positive Hermitian operator, the maximum of~\eqref{Equation: Uhlmann transition probability 2} is obtained when $U_{k+1}^\dagger U_k = V_{k+1}e^{i\sigma_k}$ for some arbitrary real constants $\sigma_k$, and is given by
\begin{equation}
    \abs{\tr(\sqrt{\sqrt{\rho_n}\rho_{n+1}\sqrt{\rho_n}})}^2\dots\abs{\tr(\sqrt{\sqrt{\rho_0}\rho_1\sqrt{\rho_0}})}^2.
    \label{Equation: Uhlmann transition probability max}
\end{equation}
Given the initial state $\sqrt{\rho_0}U^\dagger_0$, we can then write the final state as
\begin{equation}
    \sqrt{\rho_{n+1}}U^\dagger_{n+1} = \sqrt{\rho_{n+1}}V U^\dagger_0 e^{i\sigma},
\end{equation}
where
\begin{equation}
    V = V_{n+1} V_n\dots V_1, \qquad\text{and}\qquad \sigma = \sum_{k=1}^n \sigma_k.
\end{equation}
The limit of $V$ as $n\to\infty$ is a unitary operator characterising the parallel transport of the purification around the curve of density matrices.

Another way to compute $V$ at large $n$ is to use the results of the previous section, but setting $H_A=H_B=J=0$. Then at leading order~\eqref{Equation: Uhlmann transition probability 2} is equal to $\exp(-\mathcal{S})$, with $\mathcal{S}$ defined in~\eqref{Equation: generating action 2}. Since $C_k-a_k$ is anti-Hermitian, its variance
\begin{equation}
    \big(\Delta(C_k-a_k)\big)^2
\end{equation}
must be non-positive. Furthermore, assuming $\rho_k$ is invertible for all $k$, the variance vanishes if and only if $C_k-a_k=i\sigma_k$ for some arbitrary real number $\sigma_k$, and one may show that these $\sigma_k$ are the same as the previous ones. Thus, to minimise~\eqref{Equation: Uhlmann transition probability 2}, we need $C_k=a_k+i\sigma_k$, and we can solve~\eqref{Equation: C_k inverse} for $V_k$ to obtain
\begin{equation}
    V_k = \exp(-a_k + \order{\Delta t^{3/2}}).
\end{equation}
Defining $\hat{a}(t_k) = a_k/\Delta t$, in the $n\to \infty$ limit we can then formally write $V$ as a path-ordered exponential
\begin{equation}
    V=V[\rho]=\mathrm{P}\exp(-\int \hat{a}(t) \dd{t}).
\end{equation}
This is why the operators $a_k$ are important -- they describe the infinitesimal version of Uhlmann holonomy.

Usually, Uhlmann holonomy does not involve any Hamiltonians or sources. However, in the scenario described in this paper, we turn on $H_A$, $H_B$ and $J$. This effectively introduces dynamics into the Uhlmann holonomy. For this reason, and for lack of a better name, we call this `dynamical' Uhlmann holonomy.

Let us actually solve for $a_k$ now. By definition, $a_k$ obeys
\begin{equation}
    \tilde{\rho}_k a_k + a_k \tilde{\rho_k} - \sqrt{\rho_k}\delta(\sqrt{\rho_k}) + \delta(\sqrt{\rho_k})\sqrt{\rho_k} - \frac{2i\Delta t}{\hbar}\sqrt{\rho_k}(H_A+\frac12J_k\cdot O_k)\sqrt{\rho_k} = 0.
    \label{Equation: a_k equation 2}
\end{equation}
We claim that the solution to this is
\begin{equation}
    a_k = \int \dd{s} e^{-s\tilde{\rho}_k} \qty(\sqrt{\rho_k}\delta(\sqrt{\rho_k})-\delta(\sqrt{\rho_k})\sqrt{\rho_k} + \frac{2i\Delta t}{\hbar}\sqrt{\rho_k}\qty(H_A+\frac12J_k\cdot O_k)\sqrt{\rho_k}) e^{-s\tilde{\rho}_k}.
    \label{Equation: a_k solution}
\end{equation}
This integral is convergent because $\tilde{\rho}_k$ is positive, so it gives a well-defined operator. One may confirm that~\eqref{Equation: a_k solution} solves~\eqref{Equation: a_k equation 2} by direct substitution, using the fact that the anticommutator with $\tilde\rho_k$ converts into an $s$ derivative in the integral. Moreover, since the linear map
\begin{equation}
    a \mapsto \tilde{\rho}_k a + a \tilde{\rho}_k
\end{equation}
is invertible,~\eqref{Equation: a_k solution} is the unique solution to~\eqref{Equation: a_k equation 2}.

Until now, we have been using $\delta\rho_k = \order{\sqrt{\Delta t}}$, but suppose for future reference that we instead have the stronger condition $\delta\rho_k= o(\sqrt{\Delta t})$.\footnote{To remind the reader of the difference between these two equations, suppose $\epsilon$ is a small parameter. Then $f=\order{\epsilon}$ means $f/\epsilon$ is finite in the $\epsilon\to 0$ limit, whereas $f=o(\epsilon)$ means $f/\epsilon \to 0$ in the $\epsilon\to 0$ limit.} In this case, it will actually be more convenient for us to write $a$ in a slightly different form, in terms of modular flow. Given a density matrix $\rho:\mathcal{H}\to\mathcal{H}$, modular flow is a one-parameter automorphism of the algebra of operators acting on $\mathcal{H}$. It is defined by
\begin{equation}
    O \mapsto \rho^{i\alpha} O \rho^{-i\alpha},
\end{equation}
where $\alpha$ is the parameter. In Appendix~\ref{Appendix: ak in terms of modular flow}, it is shown that
\begin{equation}
    a_k = \frac1{2\pi}\int_{-\infty}^\infty\dd{y} \int_{-\infty}^\infty\dd{\alpha}e^{2i\alpha y} \rho_k^{i\alpha} \qty(\frac{1-\sech(y)}{y}\delta K_k + \frac{2i\Delta t}{\hbar}\qty(H_A + \frac12 J_k\cdot O_k) \sech(y))\rho_k^{-i\alpha} + o(\Delta t),
\end{equation}
where $K_k=-\log\rho_k$.

Note that, by assumption, $K_k = \order{1/\hbar}$. This implies $\delta K_k=\order{1/\hbar}$, since it is just the difference of two $\order{1/\hbar}$ quantities. Also, by~\eqref{Equation: classical automorphism} modular flow preserves the asymptotics of the operators it acts on, so $\rho_k^{i\alpha}\delta K_k \rho_k^{-i\alpha}=\order{1/\hbar}$. We thus have $a_k=\order{1/\hbar}$.

\subsection{Bures metric}
\label{Section: Bures metric}

The other term in $\mathcal{S}$ which we need to analyse is
\begin{equation}
    D_k = \tr(\delta(\sqrt{\rho_k})\delta(\sqrt{\rho_k}) + \frac12a_k(\sqrt{\rho_k}\delta(\sqrt{\rho_k})-\delta(\sqrt{\rho_k})\sqrt{\rho_k})).
\end{equation}
It will be useful for us to write this in a different form. First, we have
\begin{align}
    \delta(\sqrt{\rho_k}) + a_k\sqrt{\rho_k} 
    &= \int_0^\infty\dd{s}e^{-s\rho_k} \Big(\delta(\sqrt{\rho_k})\rho_k+\rho_k\delta(\sqrt{\rho_k})+\big(\sqrt{\rho_k}\delta(\sqrt{\rho_k})-\delta(\sqrt{\rho_k})\sqrt{\rho_k}\big)\sqrt{\rho_k}\Big)e^{-s\rho_k}\nonumber\\
    & \pushright{+ \order{\Delta t}\qquad}\\
    &= \sqrt{\rho_k} \int_0^\infty \dd{s} e^{-s\rho_k}\delta\rho_k e^{-s\rho_k} + \order{\Delta t}.
\end{align}
Similarly,
\begin{equation}
    \delta(\sqrt{\rho_k})-\sqrt{\rho_k}a_k = \int_0^\infty \dd{s} e^{-s\rho_k}\delta\rho_ke^{-s\rho_k}\sqrt{\rho_k}+\order{\Delta t}.
\end{equation}
Combining these, we deduce
\begin{equation}
    D_k = \tr(\delta(\sqrt{\rho_k})\delta(\sqrt{\rho_k}) + \frac12 a_k\big(\sqrt{\rho_k}\delta(\sqrt{\rho_k})-\delta(\sqrt{\rho_k})\sqrt{\rho_k}\big)) = \frac12 \tr(\delta\rho_kG_k) + \order{\Delta t^{3/2}},
    \label{Equation: Bures metric}
\end{equation}
where
\begin{equation}
    G_k = \int_0^\infty \dd{s} e^{-s\rho_k}\delta\rho_k e^{-s\rho_k}.
\end{equation}
An alternate way to define $G_k$ is as the solution of 
\begin{equation}
    \rho_k G_k + G_k \rho_k = \delta\rho_k.
\end{equation}
Note that $\frac12\tr(\delta\rho_k G_k)$ is quadratic in $\delta\rho_k$, vanishes only if $\delta \rho_k=0$, and is positive otherwise. Thus, it provides us with a metric on the space of density matrices $\mathcal{M}_A$, called the `Bures metric'~\cite{Bures,Helstrom}.

Assuming $\delta\rho_k=\order{\sqrt{\Delta t}}$, in Appendix~\ref{Appendix: Gk in terms of modular flow} it is shown that 
\begin{equation}
    G_k = R_k + \order{\Delta t},
\end{equation}
where
\begin{equation}
    R_k = \frac1{2\pi}\int_{-\infty}^\infty\dd{y} \int_{-\infty}^\infty\dd{\alpha}e^{2i\alpha y} \frac{\tanh(y)}{y} \rho_k^{i\alpha}\delta K_k \rho_k^{-i\alpha}.
\end{equation}
Therefore, we have
\begin{equation}
    D_k = \frac12\Big(\tr(\rho_k R_k) - \tr(\rho_{k-1} R_k)\Big) + \order{\Delta t^{3/2}}.
\end{equation}
In other words, $D_k$ is half of the difference between the expectation values of $R_k$ in the two states $\rho_k$ and $\rho_{k-1}$.

As in the previous section, we have $\rho_k^{i\alpha}\delta K_k\rho_k^{-i\alpha}=\order{1/\hbar}$, which implies $R_k=\order{1/\hbar}$. Taking the expectation value preserves this scaling, so the leading order part of $D_k$ is $\order{1/\hbar}$.

\subsection{Classical limit\texorpdfstring{ ($\hbar\to0$)}{}}
\label{Section: Classical limit}

Let us now consider the classical limit $\hbar\to 0$. Ignoring subleading contributions for notational convenience, we may write $\mathcal{S}$ as the sum of three terms
\begin{equation}
    \mathcal{S} = \mathcal{S}_D + \mathcal{S}_U + \mathcal{S}_J,
\end{equation}
where
\begin{align}
    \mathcal{S}_D &= \sum_{k=1}^{n+1} D_k, \\
    \mathcal{S}_U &= -\sum_{k=1}^{n+1} \big(\Delta(C_k-a_k)\big)^2, \\
    \mathcal{S}_J &= \frac{i}{\hbar} \sum_{k=1}^n\Delta t \tr(\rho_k J_k\cdot O_k). \label{Equation: S_J definition}
\end{align}
$\mathcal{S}_D$ and $\mathcal{S}_U$ are both real and non-negative, whereas $\mathcal{S}_J$ is imaginary.

Let us first consider $\mathcal{S}_D$, which is $\order{1/\hbar}$ by the results of Section~\ref{Section: Bures metric}. In the classical limit the factor of $e^{-\mathcal{S}_D}$ in the generating function will therefore be sharply peaked, and we should seek to minimise $\mathcal{S}_D$. Since $D_k=0$ if and only if $\delta\rho_k=0$, the exact minimum is attained when $\rho_k$ is constant. However, we want to be able to consider evolution between different states, so this would be too trivial. Luckily, in the simultaneous limit $\hbar,\Delta t\to 0$, a much larger class of sequences of states $\rho_k$ is allowed. Suppose for example that the curve $\rho(t)$ which $\rho_k$ approximates is differentiable. 
Then we have $\delta\rho_k=\order{\Delta t}$, so $D_k=\order{\Delta t^2}$. This implies that the overall scaling of $\mathcal{S}_D$ with respect to both $\Delta t$ and $\hbar$ is $\mathcal{S}_D=\order{\Delta t/\hbar}$. By our assumptions, $\Delta t/\hbar\to 0$ in the simultaneous limit, so $\mathcal{S}_D\to 0$. So all differentiable paths minimise $\mathcal{S}_D$. 
Actually, there will be a larger set of paths which minimise $\mathcal{S}_D$, and this set depends on the exact relationship between $\Delta t$ and $\hbar$. Let us call the set $\mathscr{C}$. Certainly $\mathscr{C}$ will contain all differentiable paths, as we have just argued. We also have $\delta\rho_k=o(\sqrt{\Delta t})$ for all paths in $\mathscr{C}$, because otherwise $\mathcal{S}_D$ grows at least as fast as $1/\hbar$.

Let us assume now that $\mathcal{S}_D$ is minimised, so that the path $\rho(t)$ is in $\mathscr{C}$. Since this means $\delta\rho_k=o(\sqrt{\Delta t})$, by the results of Section~\ref{Section: Dynamical Uhlmann holonomy} we now have $a_k=\order{1/\hbar}$. Let us write $C_k-a_k=i\sigma_k+B_k$, where $B_k$ satisfies $\tr(\rho_kB_k)=0$; this can always be made to be true by appropriately choosing $\sigma_k$. Then we have
\begin{equation}
    \mathcal{S}_U = -\sum_{k=1} \tr(\rho_k B_k^2).
\end{equation}
This is non-negative because $B_k$ must be anti-Hermitian. It is zero if and only if $B_k=0$, but $B_k$ doesn't have to be exactly at this minimum. Suppose $B_k\ne o(\Delta t/\hbar)$. Then $\mathcal{S}_U$ grows at least as fast as $\Delta t/\hbar^2$. By our assumptions, $\Delta t/\hbar^2\to \infty$ in the simultaneous limit, so $\mathcal{S}_U\to\infty$, i.e.\ $\exp(-\mathcal{S}_U)\to 0$. To avoid this, we need $B_k=o(\Delta t/\hbar)$, in which case we may write
\begin{equation}
    C_k = a_k+i\sigma_k + o\qty(\frac{\Delta t}{\hbar}).
\end{equation}
Without loss of generality we can at this point assume $\sigma_k=0$, as it is an arbitrary phase factor that cancels out in all the following calculations. Then, using~\eqref{Equation: C_k inverse}, we have
\begin{equation}
    U_{k-1}^\dagger e^{iH_B\Delta t/\hbar} U_k = \exp(a_k+o\qty(\frac{\Delta t}{\hbar})),
\end{equation}
which implies
\begin{equation}
    U_{n+1} = e^{-iH_B T/\hbar} U_0 V[\rho,J]^\dagger
\end{equation}
where
\begin{equation}
    V[\rho,J] = e^{-a_{n+1}+o(\Delta t/\hbar)}e^{-a_n+o(\Delta t/\hbar)} \dots e^{-a_1+o(\Delta t/\hbar)}.
\end{equation}

As we have pointed out, $\mathscr{C}$ contains more paths that just the differentiable ones. However, from this point on the effects of non-differentiability will not be so important, so we will take the notational shortcut of assuming that derivatives are well-defined. This is fairly standard when dealing with path integrals, but one should always keep in mind that whenever a derivative appears it is technically a formal one, and it should be understood in an appropriately regularised sense.

We can then in the $\Delta t\to 0$ limit write $V[\rho,J]$ as a path-ordered exponential
\begin{equation}
    V[\rho,J] = \mathrm{P}\exp(-\int_0^Ta(t)\dd{t} + o\qty(\frac1{\hbar})),
    \label{Equation: V rho J}
\end{equation}
where
\begin{equation}
    a(t) = \frac1{2\pi} \int_{-\infty}^\infty \dd{y}\int_{-\infty}^\infty\dd{\alpha} e^{2i\alpha y} \rho(t)^{i\alpha} \qty(\frac{1-\sech(y)}{y} \dot{K}(t) + \frac{2i}{\hbar}\qty(H_A + \frac12 J(t)\cdot O(t))\sech(y))\rho(t)^{-i\alpha}.
    \label{Equation: a differentiable}
\end{equation}
The arguments in square brackets indicate that $V[\rho,J]$ depends on the path $\rho(t)$ and the sources $J(t)$. Since $a(t)\sim 1/\hbar$, the $o(1/\hbar)$ term in~\eqref{Equation: V rho J} is subleading in the classical limit $\hbar\to 0$, so we can ignore it in the following.

Suppose the state of the system is initially $\ket{X(0)}$. To determine the classical evolution of the system after a time $T$, we need to maximise the transition probability to the final state $\ket{X(T)}$. Recall that this probability is proportional to the generating function if we set $J=0$. If the evolution of the state does not obey the conditions we have just laid out, the transition probability will be exponentially suppressed. To be precise, there must be some path $\rho(t)$ of density matrices in $\mathscr{C}$ which starts at $\rho(0)=\tr_B\ket{X(0)}\bra{X(0)}$ and ends at $\rho(T)=\tr_B\ket{X(T)}\bra{X(T)}$. Furthermore, if we write the states $\ket{X(0)}$ and $\ket{X(T)}$ as linear maps $\sqrt{\rho(0)} U(0)^\dagger$ and $\sqrt{\rho(T)}U(T)^\dagger$, then $U(0)$ and $U(T)$ must be related by the dynamical Uhlmann holonomy along $\rho(t)$ in the absence of sources, i.e.
\begin{equation}
    U(T) = e^{-iH_BT/\hbar} U(0) V[\rho,0]^\dagger.
\end{equation}
If this were not true, then by the above arguments the integrand of the generating function~\eqref{Equation: generating function 1} would always be exponentially suppressed in the classical limit, so the transition probability itself would be exponentially suppressed. However, when this condition \emph{is} obeyed, there are contributions to the generating function which are not exponentially suppressed, and integrating over these contributions gives a non-suppressed transition probability.

To summarise, in the classical limit the dominant evolution of the system takes the form
\begin{equation}
    \sqrt{\rho(0)} U^\dagger \to \sqrt{\rho(T)}\, V[\rho,0] U^\dagger e^{iH_BT/\hbar},
\end{equation}
where $\rho=\rho(t)$ is any path in $\mathscr{C}$ from $\rho(0)$ to $\rho(T)$. There is no classically dominant choice of $\rho$ in $\mathscr{C}$.

\subsection{Semiclassical correlators}
\label{Section: Semiclassical correlators}

In this section, we will compute semiclassical correlators of operators acting on the system. This means that we will assume that the `background' evolution of the system is classical, so that correlators measure quantum fluctuations about this classical background.

Let the classical background be described by a path $\bar\rho(t)$ of reduced states that begins at $\bar\rho(0)=\rho(0)$ and ends at $\bar\rho(T)=\rho(T)$, so that the initial and final states of the system are
\begin{equation}
    \sqrt{\rho_0}U_0^\dagger = \sqrt{\rho(0)}U_0^\dagger
    \qquad\text{and}\qquad
    \sqrt{\rho_{n+1}}U_{n+1}^\dagger = \sqrt{\rho(T)}V[\bar\rho,0] U_0^\dagger e^{iH_BT/\hbar}
\end{equation}
respectively. Substituting this into the generating function~\eqref{Equation: generating function polar}, one finds
\begin{equation}
    Z[X_0,X_{n+1};J] = \int\prod_{k=1}^n\Big(\dd{\rho_k}\dd{U_k} Y_k\Big)\abs{\tr(e^{-iH_BT/\hbar}U_0V[\bar\rho,0]^\dagger\sqrt{\rho_{n+1}}\sqrt{\rho_n}U^\dagger_n)}^2,
\end{equation}
Now, in the limit $\Delta t,\hbar\to 0$, all the reasoning of the previous sections still holds. One finds therefore that this integral is dominated by sequences of states $\rho_k$ which approximate a path $\rho(t)$ in $\mathscr{C}$ starting at $\rho(0)$ and ending at $\rho(T)$, and by sequences of operators $U_k$ such that
\begin{equation}
    U_n = e^{-iH_BT/\hbar} U_0 V[\rho,J]^\dagger.
\end{equation}
Substituting this in, the generating function takes the form
\begin{equation}
    Z[\bar\rho; J] = \int \Dd{\rho} \exp(-\mathcal{S}_J) \abs{\tr(\rho(T)V[\rho,J] V[\bar\rho,0]^\dagger)}^2,
    \label{Equation: generating function path integral 1}
\end{equation}
where $\mathcal{S}_J$ is defined in~\eqref{Equation: S_J definition}, and may be written in the $\Delta t\to 0$ limit as
\begin{equation}
    \mathcal{S}_J = \frac{i}{\hbar}\int_0^T \tr(\rho(t)J(t)\cdot O(t))\dd{t}.
\end{equation}
The integral in~\eqref{Equation: generating function path integral 1} is done over all paths of reduced states in $\mathscr{C}$ that begin at $\rho(0)$ and end at $\rho(T)$. We have indicated that the generating function depends on the entire background path $\bar\rho$ by including it in the square brackets on the left-hand side. Notice that the generating function now does not depend at all on $U_0$ or $H_B$.

It remains to evaluate the trace term in~\eqref{Equation: generating function path integral 1}, which can be done by using a coherent state path integral. This proceeds in the usual way. In particular, we can write
\begin{equation}
    \mel{x'}{V[\rho,J]}{x} = \lim_{\Delta t\to 0} \mel{x'}{e^{-\Delta ta(t_n)}e^{-\Delta ta(t_{-1})}\dots e^{-\Delta t a(t_1)}}{x},
\end{equation}
and inserting~\eqref{Equation: resolution} multiple times leads to
\begin{equation}
    \mel{x'}{V[\rho,J]}{x} = \lim_{\Delta t\to 0} \int \prod_{l=1}^{n-1}\dd{x_l}\prod_{k=1}^n \mel{x_k}{e^{-\Delta t a(t_k)}}{x_{k-1}},
\end{equation}
where $x_0=x$ and $x_n=x'$. For small $\Delta t$, the sequences of coherent states which contribute to this integral approximate continuous paths, and we can write it as
\begin{equation}
    \mel{x'}{V[\rho,J]}{x} = \int \Dd{x} \exp(is[x,\rho,J]/\hbar),
\end{equation}
where
\begin{equation}
    s[x,\rho,J] = \int_0^T \qty(i\hbar\braket{x}{\dot{x}} + i\hbar \mel{x}{a}{x})\dd{t}.
    \label{Equation: Uhlmann action}
\end{equation}
(Again, most paths which contribute to the path integral are non-differentiable, so the time derivative here is formal.) Using this twice, we can write
\begin{equation}
    \tr(\rho(T)V[\rho,J]V[\bar\rho,0]) = \int \Dd{x}\Dd{\bar{x}} \mel{\bar{x}(T)}{\rho(T)}{x(T)} \exp(i\big(s[x,\rho,J] - s[\bar{x},\bar{\rho},0]\big)/\hbar),
    \label{Equation: trace path integral}
\end{equation}
where the integral is done over paths of coherent states $x(t),\bar{x}(t)$ which obey $x(0)=\bar{x}(0)$.

We can now substitute this into the generating function~\eqref{Equation: generating function path integral 1}. Actually,~\eqref{Equation: trace path integral} appears twice in~\eqref{Equation: generating function path integral 1}, once as a complex conjugate. We can deal with this by doubling the degrees of freedom $x\to x_L,x_R$ and $\bar{x}\to\bar{x}_L,\bar{x}_R$. We end up with
\begin{equation}
    Z[\rho(0),\rho(T);J] = \int \Dd{\rho}\Dd{x_L}\Dd{x_R}\Dd{\bar{x}_L}\Dd{\bar{x}_R} p\, \exp(iS/\hbar),
    \label{Equation: generating function path integral 2}
\end{equation}
where the overall action is
\begin{equation}
    S = s[x_L,\rho,J] - s[x_R,\rho,J] - s[\bar{x}_L,\bar\rho,0] + s[\bar{x}_R,\bar\rho,0] - \int_0^T \tr(\rho J\cdot O) \dd{t},
    \label{Equation: generating function action}
\end{equation}
and 
\begin{equation}
    p = \mel{\bar{x}_L(T)}{\rho(T)}{x_L(T)}\mel{x_R(T)}{\rho(T)}{\bar{x}_R(T)}.
\end{equation}
Since $\rho= e^{-\order{1/\hbar}}$, $p$ is sharply peaked in the classical limit. Therefore, its effect is just to set some boundary conditions on $x_{L,R},\bar{x}_{L,R}$ at $t=T$.

Since the Berry connection $i\braket{x}{\dot{x}}$ is assumed to be $\order{1/\hbar}$, and we have shown $a$ is $\order{1/\hbar}$, the action~\eqref{Equation: generating function action} is $\order{1}$, so we can treat it as a classical action. Thus, we can compute all semiclassical correlators by applying methods of stationary phase, and other similar tools, to the formula
\begin{equation}
    \expval{O(t_m)\dots O(t_1)} = \frac{(i\hbar)^m}{Z[\bar\rho;0]}\left.\fdv{J(t_m)}\dots\fdv{J(t_1)} Z[\bar\rho;J]\right|_{J=0}.
    \label{Equation: holographic correlator}
\end{equation}

\section{Emergent holography}
\label{Section: Emergent holography}

We will now show that~\eqref{Equation: generating function path integral 2} is a holographic generating function in disguise.

Let us define a set of new states $\ket{x,\alpha}$ by acting on the coherent states $\ket{x}$ with modular flow:
\begin{equation}
    \ket{x,\alpha} = \rho^{-i\alpha}\ket{x}.
\end{equation}
These states clearly obey the `modular Schr\"odinger equation'
\begin{equation}
    \partial_\alpha\ket{x,\alpha} = iK \ket{x,\alpha},
\end{equation}
which implies that
\begin{align}
    \mel{x,\alpha}{\dot{K}}{x,\alpha} &= \bra{x,\alpha}\qty(\pdv{t}(K\ket{x,\alpha}) - K\pdv{t}\ket{x,\alpha}) \\
                                      &= -i\bra{x,\alpha}\pdv{t}\pdv{\alpha}\ket{x,\alpha} -i\pdv{\alpha}\bra{x,\alpha}\pdv{t}\ket{x,\alpha} \\
                                      &= -i\pdv{\alpha}\qty(\bra{x,\alpha}\pdv{t}\ket{x,\alpha}).
\end{align}
Using this we can write the diagonal coherent state elements of~\eqref{Equation: a differentiable} as
\begin{align}
    \mel{x}{a}{x} &= \frac1{2\pi}\int_{-\infty}^\infty\dd{y}\int_{-\infty}^\infty\dd{\alpha}e^{2i\alpha y}\bra{x,\alpha}\qty(\frac{1-\sech(y)}{y}\dot{K} + \frac{2i}{\hbar}\qty(H_A+\frac12J\cdot O)\sech(y))\ket{x,\alpha} \\
                  &= \frac1{\pi}\int_{-\infty}^\infty\dd{y}\int_{-\infty}^\infty\dd{\alpha}e^{2i\alpha y}\bigg(\bra{x,\alpha}\pdv{t}\ket{x,\alpha}\qty(\sech(y)-1) \nonumber\\
                  &\qquad\qquad\qquad\qquad\qquad\qquad\qquad\qquad\qquad + \frac{i}{\hbar}\bra{x,\alpha}\qty(H_A+\frac12 J\cdot O)\ket{x,\alpha}\sech(y)\bigg) \\
                  &= -\bra{x}\pdv{t}\ket{x} + \int_{-\infty}^\infty \dd{\alpha}\sech(\pi\alpha)\qty(\bra{x,\alpha}\pdv{t}\ket{x,\alpha} + \frac{i}{\hbar}\bra{x,\alpha}\qty(H_A+\frac12 J\cdot O)\ket{x,\alpha}).
\end{align}
In the second line we integrated by parts with respect to $\alpha$ on the $\dot{K}$ term, and in the third line we used the well-known Fourier transforms
\begin{equation}
    \int_{-\infty}^\infty e^{2i\alpha y} \dd{y} = \pi \delta(\alpha)
    \qquad\text{and}\qquad
    \int_{-\infty}^\infty e^{2i\alpha y} \sech(y) = \pi \sech(\pi\alpha).
\end{equation}
Substituting this into~\eqref{Equation: Uhlmann action}, we find
\begin{align}
    s[x,\rho,J] &= \int_0^T\dd{t}\qty(i\hbar\bra{x}\pdv{t}\ket{x} + i\hbar\mel{x}{a}{x}) \\
                &= \int_0^T\dd{t}\int_{-\infty}^{\infty}\dd{\alpha} \sech(\pi\alpha) \qty(i\hbar\bra{x,\alpha}\pdv{t}\ket{x,\alpha} - \bra{x,\alpha}\qty(H_A+\frac12 J\cdot O)\ket{x,\alpha}).
                \label{Equation: holographic Uhlmann action}
\end{align}

Recall the action~\eqref{Equation: isolated action} for the evolution of the coherent states in the isolated case:
\begin{equation}
    S = \int_0^T\dd{t}\qty(i\hbar\bra{x}\pdv{t}\ket{x} - \mel{x}{H}{x}).
    \label{Equation: isolated action 2}
\end{equation}
In a very immediate sense, we see that~\eqref{Equation: holographic Uhlmann action} has one more dimension than~\eqref{Equation: isolated action 2}. This dimension is parametrised by $\alpha$, i.e.\ it is generated by modular flow. 

In the isolated case, we can recognise~\eqref{Equation: isolated action 2} as a Hamiltonian action with symplectic form
\begin{equation}
    \omega = \lim_{\hbar\to 0}i\hbar \dd{\bra{x}}\wedge\dd{\ket{x}}.
\end{equation}
and Hamiltonian function
\begin{equation}
    h(x) = \lim_{\hbar\to 0}\mel{x}{H}{x}.
\end{equation}
By choosing canonical coordinates $p_i,q_i$ on the space of coherent states, we can write this as
\begin{equation}
    \omega = \sum_i \dd{q_i}\wedge\dd{p_i} \qquad\text{and}\qquad h=h(p_i,q_i).
\end{equation}
These canonical coordinates $p_i,q_i$ represent the classical degrees of freedom.

In the classical limit, modular flow reduces to a kind of classical evolution on the space of coherent states. To see this, note that 
\begin{equation}
    \ket{x,\alpha} = \rho^{-i\alpha} \ket{x} = e^{i\alpha K}\ket{x}
\end{equation}
can be computed with a coherent state path integral by replacing $H\to -\hbar K$ in the usual transition amplitude. Then, because of our assumption that the modular Hamiltonian obeys $K=\order{1/\hbar}$, there will be a dominant path in the classical limit. So $\ket{x,\alpha}$ corresponds to a single coherent state in the classical limit; let us write its canonical coordinates as $p_i(\alpha),q_i(\alpha)$.

The action~\eqref{Equation: holographic Uhlmann action} is also a Hamiltonian action. The symplectic form is
\begin{equation}
    \Omega = \lim_{\hbar\to 0}i\hbar \int_{-\infty}^\infty\dd{\alpha} \sech(\pi\alpha) \, \dd{\bra{x,\alpha}}\wedge\dd{\ket{x,\alpha}}.
\end{equation}
In terms of the canonical coordinates this is
\begin{equation}
    \Omega = \int_{-\infty}^\infty\dd{\alpha}\sech(\pi\alpha) \,\sum_i\dd{q_i(\alpha)}\wedge\dd{p_i(\alpha)}.
    \label{Equation: holographic symplectic form}
\end{equation}
The Hamiltonian function can be found by setting the sources to zero; it is
\begin{equation}
    H(x,\rho) = \lim_{\hbar\to 0} \int_{-\infty}^\infty\dd{\alpha}\sech(\pi\alpha)\mel{x,\alpha}{H_A}{x,\alpha}.
\end{equation}
If the $H_A$ here is the same operator as the $H$ in the isolated case, then we have in terms of the canonical coordinates
\begin{equation}
    H(x,\rho) = \int_{-\infty}^\infty\dd{\alpha}\sech(\pi\alpha)\, h\big(p_i(\alpha),q_i(\alpha)\big).
    \label{Equation: holographic hamiltonian function}
\end{equation}
\eqref{Equation: holographic symplectic form} and~\eqref{Equation: holographic hamiltonian function} provide another perspective on the emergent holographic dimension generated by modular flow. We clearly see that there is an additional dimension's worth of classical degrees of freedom labelled by $\alpha$.

On the other hand, the degrees of freedom $p(\alpha_1),q(\alpha_1)$ and $p(\alpha_2),q(\alpha_2)$ for $\alpha_1\ne\alpha_2$ are not actually independent of one another, but are related by the modular Schr\"odinger equation. However, this kind of spacelike constraint on the degrees of freedom is actually something we should expect to happen in the bulk theory, because the bulk theory should have some gauge symmetries. Thus we can view the modular Schr\"odinger equation as a bulk gauge constraint.

The action~\eqref{Equation: holographic Uhlmann action} appears four times in the overall action for the generating function~\eqref{Equation: generating function path integral 2}. Two of these just account for the background evolution $\bar\rho$ without sources, and should be viewed as counterterms. So really there are two different sourced holographic `sectors' to the path integral, labelled by $L$ and $R$. This factor of 2 accounts for the factor of $\frac12$ in front of the sources in~\eqref{Equation: holographic Uhlmann action}.

\begin{figure}
    \centering
    \begin{tikzpicture}[scale=0.9]
        \begin{scope}[red,semithick,%
            decoration={markings,
            mark=at position 0.1 with {\coordinate(A1);},
            mark=at position 0.5 with {\coordinate(B1);},
            mark=at position 0.9 with {\coordinate(C1);}
            }]
            \path[postaction={decorate}] (1,0) .. controls (0,3) and (2,6) .. (1,8);
            \path (A1) 
                --+ (0,-1) coordinate (A2) 
                --+ (5,-2) coordinate (A3)
                --+ (6,-2) coordinate (A4);
            \path (A1) 
                --+ (0,1) coordinate (A2b) 
                --+ (5,2) coordinate (A3b)
                --+ (6,2) coordinate (A4b);
            \path (B1) 
                --+ (0,-1) coordinate (B2) 
                --+ (5,-2) coordinate (B3)
                --+ (6,-2) coordinate (B4);
            \path (B1) 
                --+ (0,1) coordinate (B2b) 
                --+ (5,2) coordinate (B3b)
                --+ (6,2) coordinate (B4b);
            \path (C1) 
                --+ (0,-1) coordinate (C2) 
                --+ (5,-2) coordinate (C3)
                --+ (6,-2) coordinate (C4);
            \path (C1) 
                --+ (0,1) coordinate (C2b) 
                --+ (5,2) coordinate (C3b)
                --+ (6,2) coordinate (C4b);

            \fill[red!10] (0,1.5) -- (A1) .. controls (A2b) and (A3b) .. (A4b) 
                .. controls (B4b) .. (C4b)
                .. controls (C3b) and (C2b) .. (C1) -- (0,6.5);

            \fill[red!30,opacity=0.8] (0,1.5) -- (A1) .. controls (A2) and (A3) .. (A4) 
                .. controls (B4) .. (C4)
                .. controls (C3) and (C2) .. (C1) -- (0,6.5);

            \draw[dashed,red!70] (A1) .. controls (A2b) and (A3b) .. (A4b);
            \draw (A1) .. controls (A2) and (A3) .. (A4);
            \draw[dashed,red!70] (B1) .. controls (B2b) and (B3b) .. (B4b);
            \draw (B1) .. controls (B2) and (B3) .. (B4);
            \draw[dashed,red!70] (C1) .. controls (C2b) and (C3b) .. (C4b);
            \draw (C1) .. controls (C2) and (C3) .. (C4);
        \end{scope}
        \begin{scope}[red,semithick,%
            decoration={markings,
            mark=at position 0.1 with {\coordinate(A1);},
            mark=at position 0.5 with {\coordinate(B1);},
            mark=at position 0.9 with {\coordinate(C1);}
            }]
            \path[postaction={decorate}] (-1,0) .. controls (-2,3) and (0,6) .. (-1,8);
            \path (A1) 
                --+ (0,-1) coordinate (A2) 
                --+ (-5,-2) coordinate (A3)
                --+ (-6,-2) coordinate (A4);
            \path (A1) 
                --+ (0,1) coordinate (A2b) 
                --+ (-5,2) coordinate (A3b)
                --+ (-6,2) coordinate (A4b);
            \path (B1) 
                --+ (0,-1) coordinate (B2) 
                --+ (-5,-2) coordinate (B3)
                --+ (-6,-2) coordinate (B4);
            \path (B1) 
                --+ (0,1) coordinate (B2b) 
                --+ (-5,2) coordinate (B3b)
                --+ (-6,2) coordinate (B4b);
            \path (C1) 
                --+ (0,-1) coordinate (C2) 
                --+ (-5,-2) coordinate (C3)
                --+ (-6,-2) coordinate (C4);
            \path (C1) 
                --+ (0,1) coordinate (C2b) 
                --+ (-5,2) coordinate (C3b)
                --+ (-6,2) coordinate (C4b);

            \fill[red!10] (0,1.5) -- (A1) .. controls (A2b) and (A3b) .. (A4b) 
                .. controls (B4b) .. (C4b)
                .. controls (C3b) and (C2b) .. (C1) -- (0,6.5);

            \fill[red!30,opacity=0.8] (0,1.5) -- (A1) .. controls (A2) and (A3) .. (A4) 
                .. controls (B4) .. (C4)
                .. controls (C3) and (C2) .. (C1) -- (0,6.5);

            \draw[dashed,red!70] (A1) .. controls (A2b) and (A3b) .. (A4b);
            \draw (A1) .. controls (A2) and (A3) .. (A4);
            \draw[dashed,red!70] (B1) .. controls (B2b) and (B3b) .. (B4b);
            \draw (B1) .. controls (B2) and (B3) .. (B4);
            \draw[dashed,red!70] (C1) .. controls (C2b) and (C3b) .. (C4b);
            \draw (C1) .. controls (C2) and (C3) .. (C4);
        \end{scope}
        \fill[blue!20] (-1,0) .. controls (-2,3) and (0,6) .. (-1,8)
            -- (1,8) .. controls (2,6) and (0,3) .. (1,0) -- cycle;

        \draw[line width=3pt,blue,->] (1/6,-0.5) -- (0,0) .. controls (-1,3) and (1,6) .. (0,8) -- (-1/4,8.5);
        \draw[semithick,blue] (1,0) .. controls (0,3) and (2,6) .. (1,8);
        \draw[semithick,blue] (-1,0) .. controls (-2,3) and (0,6) .. (-1,8);

        \fill[blue] (0,4.4) circle (0.15);
        \node[above left] at (0,4.4) {\huge$\rho$};

        \draw[->,very thick] (-4,3.5) -- (-3.9,4.5) node[above] {\Large$t$};
        \draw[->,very thick] (-4,3.5) -- (-5,3.2) node[left] {\Large$\alpha$};
        \fill (-4,3.5) circle (0.08);

        \fill[red] (-5,0.5) circle (0.15);
        \node[above right] at (-5,0.5) {\LARGE$\ket{x_L,\alpha}$};

        \draw[->,very thick] (4,0.7) -- (4,1.7) node[above] {\Large$t$};
        \draw[->,very thick] (4,0.7) -- (3,1) node[left] {\Large$\alpha$};
        \fill (4,0.7) circle (0.08);

        \fill[red] (4.5,7) circle (0.15);
        \node[above right] at (4.5,7) {\LARGE$\ket{x_R,\alpha}$};

        \node[left] at (-7.5,4) {\Huge$L$};
        \node[right] at (7.5,4) {\Huge$R$};
    \end{tikzpicture}
    \caption{The emergent bulk spacetime has two sides $L$ and $R$, which are both generated by the modular flow of a single evolving density matrix $\rho$.}
    \label{Figure: emergent theory}
\end{figure}
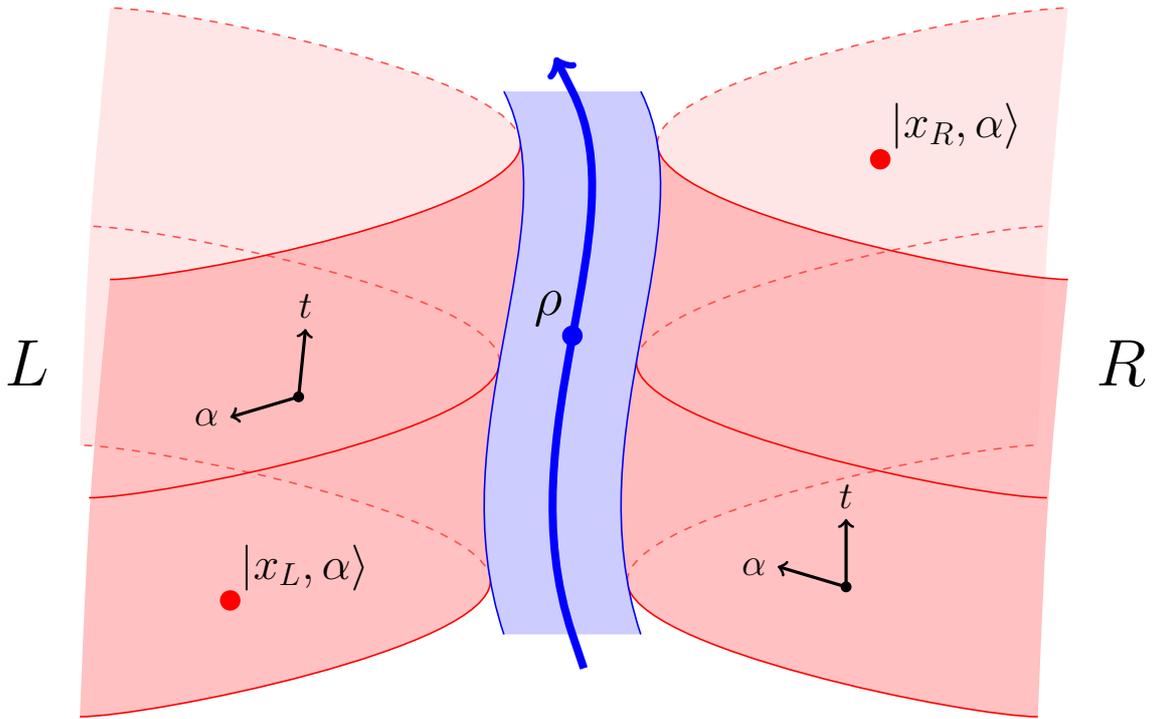

Let us summarise exactly what the emergent holographic theory looks like.
\begin{itemize}
    \item There is a fixed path $\bar{\rho}$ of density matrices which represents the classical `background' around which we are considering fluctuations.
    \item There is another path $\rho$ of density matrices which is not fixed. Instead, we integrate over $\rho\in\mathscr{C}$ in the path integral, subject to the constraints $\rho(0)=\bar{\rho}(0)$ and $\rho(T)=\bar{\rho}(T)$.
    \item There are two `sides', which we have been labelling left and right, $L$ and $R$. 
    \item On each side there are two families of states $\ket{x,\alpha}$ and $\ket{\bar{x},\alpha}$. The first family represents the holographic bulk generated by the modular flow of the fluctuating density matrix $\rho$ in the presence of sources $J$, while the second family represents the holographic bulk generated by the modular flow of the background density matrix $\bar\rho$.
    \item On each side we integrate over $x$ and $\bar{x}$, subject to the constraint $x(0)=\bar{x}(0)$. Since $\rho(0) = \bar\rho(0)$, this more or less says that at $t=0$ the state of the holographic bulk contains no fluctuations around the background.
\end{itemize}
The emergent bulk spacetime is depicted in Figure~\ref{Figure: emergent theory}.

We can measure the extra dimension by computing correlators, i.e.\ by taking derivatives of the generating function with respect to the sources $J$ and using the formula~\eqref{Equation: holographic correlator}. One might be concerned that the extra dimension is trivial, in the sense that the degrees of freedom $p_i(\alpha),q_i(\alpha)$ for different values of $\alpha$ might evolve more or less independently of one another. However, this seems very unlikely, given that the density matrix $\rho$ is a fluctuating object. Since $\rho$ is what determines the relationship between the degrees of freedom at different values of $\alpha$, this relationship must also fluctuate. Moreover, these fluctuations contribute to the action, and so should be measurable.

Another concern may be that the operators $O$ in the action~\eqref{Equation: holographic Uhlmann action} must be applied uniformly for all $\alpha$. This could mean we won't be able to measure individual degrees of freedom in the extra dimension. However, we should point out that if we are allowed to define state-dependent observables then we \emph{can} insert operators at fixed points in the emergent dimension. Indeed, by the methods of Appendix~\ref{Appendix: ak in terms of modular flow}, the part of $a$ involving the sources, i.e.
\begin{equation}
    a_J=\frac{i}{2\hbar}\int_{-\infty}^\infty\dd{\alpha}\sech(\pi\alpha)\rho^{i\alpha}J\cdot O\rho^{-i\alpha},
    \label{Equation: aJ}
\end{equation}
is the solution to
\begin{equation}
    \rho a_J + a_J \rho = \frac{i}{\hbar}\sqrt{\rho} J\cdot O \sqrt{\rho}.
    \label{Equation: aJ satisfies}
\end{equation}
If we want to insert an operator at $\alpha=\alpha_0$, then we need to find a $J\cdot O$ such that 
\begin{equation}
    a_J = \frac{i}{2\hbar}\rho^{i\alpha_0} j(\alpha_0)\cdot O\rho^{-i\alpha_0},
\end{equation}
where $j(\alpha_0)$ has the interpretation of a local source at $\alpha_0$. Substituting this into~\eqref{Equation: aJ satisfies} and rearranging, we find
\begin{equation}
    J\cdot O = j(\alpha_0)\cdot O(\alpha_0),
    \label{Equation: local source}
\end{equation}
where
\begin{equation}
    O(\alpha_0) = \frac12\rho^{i\alpha_0} \qty(\rho^{1/2}O\rho^{-1/2} + \rho^{-1/2}O\rho^{1/2})\rho^{-i\alpha_0}.
    \label{Equation: local operator}
\end{equation}
If we insert $O(\alpha_0)$ into the correlator, this will result in an insertion of $O$ at $\alpha=\alpha_0$ in the holographic bulk. To be precise, we will get an action of the form
\begin{multline}
    s[x,\rho,J] = \int_0^T\dd{t}\Bigg[\int_{-\infty}^{\infty}\dd{\alpha} \sech(\pi\alpha) \qty(i\hbar\bra{x,\alpha}\pdv{t}\ket{x,\alpha} - \bra{x,\alpha}H_A\ket{x,\alpha})\\
        - \frac{i}{2\hbar}\mel{x,\alpha_0}{j(\alpha_0)\cdot O}{x,\alpha_0}\Bigg].
\end{multline}

Of course, we do not have access to the full quantum state $\rho$ during the evolution, and even if we did it would not be consistent with the postulates of quantum mechanics to define these kinds of non-linear observables. However, what we \emph{do} have access to is the fixed background evolution $\bar\rho$. If we replace $\rho$ by $\bar\rho$ in~\eqref{Equation: local operator}, we will get a kind of local operator which is fully consistent with quantum mechanics. We should also point out that in a gravitational theory there are no local gauge-invariant bulk observables. A common way to deal with this is to define observables with respect to some fixed background, which is exactly what we would be doing in this case. It is interesting that an often-claimed desirable feature of a fundamental theory of quantum gravity is background independence, whereas here the background is built into the mechanism.

\section{Discussion}
\label{Section: Discussion}
So, starting with a highly entangled system undergoing frequent decoherence, we have obtained a genuine, non-trivial, emergent holographic theory. Let us now discuss some potential future directions.

The most obvious thing to do first would be to try to relax some of the assumptions laid out in the Introduction. In the paper we assumed that $\mathcal{H}_A=\mathcal{H}_B$, but it should not be too difficult to consider the case where the subsystems are different. Likewise, it should be straightforward to generalise the calculations to include correlators of operators applied to both $A$ and $B$. A potentially more difficult and more interesting problem would be to allow an interaction term in the system Hamiltonian~\eqref{Equation: system Hamiltonian}.

Next, although the mechanism clearly gives a holographic bulk, it is not completely obvious that the bulk theory includes gravity, and there are many questions that one could ask about this. For example, is there a fluctuating bulk metric, and do the bulk fields respect a bulk diffeomorphism gauge symmetry? One aspect of the bulk spacetime that clearly hints at gravity is the sense in which it resembles a wormhole. Indeed, it has two sides $L$ and $R$, and wormholes often have hyperbolic functions appearing in their geometry, which the factor of $\sech(\pi\alpha)$ in~\eqref{Equation: holographic Uhlmann action} seems to account for (although obviously this could just be a coincidence). A gravitational theory should also involve some kind of sum over topologies, and this could be explored by going beyond the bipartite entanglement considered in this paper. In particular, we could expand to some large number of subsystems $A,B,C,\dots$, and consider pointer states with a large amount of multipartite entanglement among these subsystems. It may be reasonable to consider pointer states formed from tensor networks. If we sum over tensor network pointer states with different underlying graphs, this would then give a sum over topologies. Of course, to extend the techniques in this paper to multipartite entanglement, one would have to have some kind of multipartite generalisation of Uhlmann holonomy. This could be a worthwhile topic to explore in its own right.

There are several approximations which are made in the paper, and it would be worthwhile to investigate corrections to these approximations. This includes the simple model of decoherence we have used -- it would be interesting to see if a more complicated model leads to significant changes to our results. There are quantum corrections to the $\hbar\to 0$ limit that should be computed. Indeed, one should figure out whether the mechanism can be made to work at all, if we are away from this classical limit, which has seemed quite essential to our derivation. There will also be corrections to the $\Delta t \to 0$ limit. In reality, $\Delta t$ is not arbitrarily small, but represents a lower limit on the temporal resolution of the theory. In the gravitational context, it is tempting to interpret this lower limit as being associated with some kind of underlying discreteness of the bulk spacetime, so that $\Delta t$ is the Planck time. It would be very interesting to see how far this interpretation goes.

One question worth asking is: what is the Hilbert space of the bulk theory? From a certain point of view, the answer is trivial. The Hilbert space must be $\mathcal{H}_A\otimes\mathcal{H}_B\otimes\mathcal{H}_{\text{E}}$, where $\mathcal{H}_{\text{E}}$ is the Hilbert space of the environment, because that is what we started with in our construction. However, suppose we didn't know that this was the starting point, and we were unaware of the mechanism described in this paper. Instead, suppose that after many experiments we had empirically deduced that physics is well-described by the generating function~\eqref{Equation: generating function path integral 2}, with each side of spacetime described by an action of the form~\eqref{Equation: holographic Uhlmann action}. Let's just focus on one side. What Hilbert space would be consistent with this action?

Using the fact that at each moment in time we have a state $\ket{x,\alpha}\in\mathcal{H}_A$ for each $\alpha\in\RR$, a reasonable first guess at an answer would be
\begin{equation}
    \hat{\mathcal{H}}_{\text{bulk}} = \bigotimes_{\alpha\in\RR}\mathcal{H}^\alpha_A.
    \label{Equation: bulk hilbert space 1}
\end{equation}
Here we have attached a label $\alpha$ to each copy of $\mathcal{H}_A$. Suppose we have two states in $\hat{\mathcal{H}}_{\text{bulk}}$:
\begin{align}
    \ket{\Phi} &= \bigotimes_{\alpha\in\RR} \ket{\Phi^\alpha},\\
    \ket*{\tilde\Phi} &= \bigotimes_{\alpha\in\RR} \ket*{\tilde\Phi^\alpha},
\end{align}
where $\ket{\Phi^\alpha},\ket*{\tilde\Phi^\alpha}\in\mathcal{H}_A^\alpha$. Then we define the inner product of these states as
\begin{equation}
    \braket*{\Phi}{\tilde\Phi} = \int_{-\infty}^{\infty} \dd{\alpha}\sech(\pi\alpha)\braket*{\Phi^\alpha}{\tilde\Phi^\alpha}.
    \label{Equation: bulk inner product}
\end{equation}
The factor of $\sech(\pi\alpha)$ here ensures that the right factor of $\sech(\pi\alpha)$ would appear in the action of a path integral formed from these states. It also means that if $\ket{\Phi^\alpha}$ is normalised for all $\alpha$, then so is $\ket{\Phi}$, since 
\begin{equation}
    \int_{-\infty}^{\infty}\dd{\alpha}\sech(\pi\alpha) = 1.
\end{equation}
More generally, if $\braket{\Phi^\alpha}{\Phi^\alpha}$ grows less quickly than $e^{\pi\abs{\alpha}}$ as $\abs{\alpha}\to \infty$, then $\ket{\Phi}$ will be a normalisable state. 

It seems however that~\eqref{Equation: bulk hilbert space 1} contains too many states. Indeed, the bulk states that we have constructed obey the modular Schr\"odinger equation
\begin{equation}
    \partial_\alpha \ket{\Phi^\alpha} = iK(X) \ket{\Phi^\alpha},
    \label{Equation: modular Schrodinger}
\end{equation}
where $K(X)$ is the modular Hamiltonian of some pointer state $X\in\mathcal{M}$. Clearly most states in~\eqref{Equation: bulk hilbert space 1} do not obey this equation. It is tempting to try to define the bulk Hilbert space so that we restrict to states for which the equation is obeyed, writing something like
\begin{equation}
    \mathcal{H}_{\text{bulk}} \stackrel{?}{=} \qty{ \ket{\Phi}\in\hat{\mathcal{H}}_{\text{bulk}} : \partial_\alpha \ket{\Phi^\alpha} = iK(X)\ket{\Phi^\alpha}\text{ for some }X\in\mathcal{M} }.
    \label{Equation: true H bulk?}
\end{equation}
But this cannot possibly work in general. The reason is that if two states obey~\eqref{Equation: modular Schrodinger} for different pointer states $X_1,X_2$, then their sum in general will not obey~\eqref{Equation: modular Schrodinger} for any pointer state. So the $\mathcal{H}_{\text{bulk}}$ in~\eqref{Equation: true H bulk?} would not be a vector space. What we could do instead is fix $X$ to start with, and define
\begin{equation}
    \mathcal{H}_{\text{bulk}}[X] = \qty{ \ket{\Phi}\in\hat{\mathcal{H}}_{\text{bulk}} : \partial_\alpha \ket{\Phi^\alpha} = iK(X)\ket{\Phi^\alpha} }.
\end{equation}
Because $X$ is fixed,~\eqref{Equation: modular Schrodinger} now \emph{is} preserved if we add two states in $\mathcal{H}_{\text{bulk}}(X)$. It seems that the correct thing to do may be to define the bulk Hilbert space as the sum of these spaces over all $X$, i.e.
\begin{equation}
    \mathcal{H}_{\text{bulk}} = \sum_{X\in\mathcal{M}} \mathcal{H}_{\text{bulk}}[X] = \operatorname{span} \qty{ \ket{\Phi}\in\hat{\mathcal{H}}_{\text{bulk}} : \partial_\alpha \ket{\Phi^\alpha} = iK(X)\ket{\Phi^\alpha}\text{ for some }X\in\mathcal{M} }.
\end{equation}
This $\mathcal{H}_{\text{bulk}}$ will in general be a proper subspace of $\hat{\mathcal{H}}_{\text{bulk}}$. It will also in general contain states which do not obey the modular Schr\"odinger equation -- the interpretation of these states is a puzzle which needs solving.

It would be interesting to see if one could make a connection with quantum error correction approaches to holography by interpreting $\mathcal{H}_{\text{bulk}}$ as a code subspace inside of $\hat{\mathcal{H}}_{\text{bulk}}$. Note that the usual technological purpose of quantum error correction is to protect a system from decoherence with the environment. Here we seem to have the opposite: the decohering process is what causes the code subspace to be favoured.

Finally, although this paper has mainly taken a phenomenological point of view, we will at some point have to provide a fundamental theory which fulfils the assumptions we have made. It would be interesting to see whether the string theoretic setups that lead to AdS/CFT do this. But it may be possible to find a much simpler toy theory with the right properties.

\section*{Acknowledgements}
\addcontentsline{toc}{section}{\protect\numberline{}Acknowledgements}
Thank you to 
William Donnelly, Tanguy Grall, Kelley Kirklin, Alex Maloney, Jo\~ao Melo, Malcolm Perry, G\'abor S\'arosi, Jakub Supe\l{} and Aron Wall
for useful discussions and comments. This work was supported by a grant from STFC. Thanks also to my upstairs neighbour, whose percussive complaints about my piano playing forced me to focus on finishing the paper. These results were obtained during the pandemic lockdown, but it would probably be in poor taste to thank COVID-19.

\appendix

\section{Generating function for a decohering system}
\label{Appendix: Generating function for decohering system}

\subsection{Correlators in open systems}
\label{Section: Correlators in open systems}

Consider an open system with Hilbert space $\mathcal{H}$, coupled to an environment with Hilbert space $\mathcal{H}_{\text{E}}$, evolving with a combined Hamiltonian $\mathbf{H}$. Suppose we are completely ignorant of the state of the environment. Then we could describe the environment in terms of the maximally mixed density matrix
\begin{equation}
    \rho_{0,\text{E}} = \frac{I_{\text{E}}}{\dim(\mathcal{H}_{\text{E}})}.
\end{equation}
However, here it will be more useful to think of the environment as being in a random normalised pure state $\ket{\phi}$, distributed according to some probability measure $\dd{\phi}$. This is equivalent to the density matrix description. The probability measure must be invariant under $\ket{\phi}\to V\ket{\phi}$ for any unitary operator $V$, since
\begin{equation}
    \mel{\phi}{\rho_{0,\text{E}}}{\phi} =
    \mel{\phi}{V^\dagger\rho_{0,\text{E}}V}{\phi}.
\end{equation}
There is essentially only one such probability measure, induced from the invariant measure $\dd{V}$ on $U(\mathcal{H}_{\text{E}})$ by setting $\ket{\phi}=V\ket{\Omega}$ for some fixed $\ket{\Omega}$. This measure obeys the useful formula
\begin{equation}
    \int\dd{\phi}\ket{\phi}\bra{\phi} = \rho_{0,\text{E}}.
    \label{Equation: environment resolution}
\end{equation}

Suppose the combined system-environment initial and final states $\ket{\Psi},\ket{\Psi'}$ take the form
\begin{equation}
    \ket{\Psi} = \ket{\psi}\otimes\ket{\phi}, \qquad
    \ket{\Psi'} = \ket{\psi'}\otimes\ket{\phi'},
\end{equation}
where $\ket{\phi},\ket{\phi'} \in \mathcal{H}_{\text{E}}$ are distributed according to the above probability measure. These states involve no entanglement between $\mathcal{H}$ and $\mathcal{H}_{\text{E}}$, i.e.\ no correlations between the system and environment.

For a given $\ket{\phi},\ket{\phi'}$, the transition amplitude for the evolution between these states after a time $T$ is
\begin{equation}
    \mel{\Psi'}{e^{-i\mathbf{H}T/\hbar}}{\Psi'} = \bra{\psi'}\otimes \bra{\phi'}e^{-i\mathbf{H}t/\hbar}\ket{\psi}\otimes \ket{\phi}.
\end{equation}
If we average this over $\ket{\phi}$ and $\ket{\phi'}$, it should be clear that we get 0, because there is a uniform integration over an arbitrary phase. Thus, the transition amplitude should not be viewed as a measurable quantity.

The story is different for the transition probability. For given $\ket{\phi},\ket{\phi'}$, this is
\begin{align}
    \mathbb{P}(\psi\to\psi'|\phi,\phi') &= \abs{\mel{\Psi'}{e^{-i\mathbf{H}t/\hbar}}{\Psi'}}^2 \\
    &= \bra{\psi'}\otimes \bra{\phi'}e^{-i\mathbf{H}t/\hbar}\ket{\psi}\otimes \ket{\phi}
    \bra{\psi}\otimes \bra{\phi}e^{i\mathbf{H}t/\hbar}\ket{\psi'}\otimes \ket{\phi'}\\
    &= \tr(\big(\rho'\otimes\ket{\phi'}\bra{\phi'}\big)e^{-i\mathbf{H}t/\hbar}\big(\rho\otimes\ket{\phi}\bra{\phi}\big)e^{i\mathbf{H}t/\hbar}).
\end{align}
where $\rho=\ket{\psi}\bra{\psi}$ and $\rho'=\ket{\psi'}\bra{\psi'}$. The notation $\mathbb{P}(\psi\to\psi'|\phi,\phi')$ is meant to emphasise that this is a conditional probability -- it is the probability of a transition $\psi\to\psi'$, conditional on the environment states being $\phi$ and $\phi'$. Using~\eqref{Equation: environment resolution} the overall transition probability is then given by
\begin{align}
    \mathbb{P}(\psi\to\psi') &= \sum_{\phi,\phi'} \mathbb{P}(\psi\to\psi'|\phi,\phi')\mathbb{P}(\phi)\mathbb{P}(\phi') \\
                             &= \int \dd{\phi}\int\dd{\phi'}\tr(\big(\rho'\otimes\ket{\phi'}\bra{\phi'}\big)e^{-i\mathbf{H}t/\hbar}\big(\rho\otimes\ket{\phi}\bra{\phi}\big)e^{i\mathbf{H}t/\hbar}) \\
                             &= \frac1{(\dim(\mathcal{H}_{\text{E}}))^2}\tr(\qty(\rho'\otimes I_{\text{E}})\,e^{-i\mathbf{H}t/\hbar}\,\qty(\rho\otimes I_{\text{E}})\,e^{i\mathbf{H}t/\hbar}).
\end{align}
Note that this is identical to the transition probability between the density matrices $\rho\otimes \rho_{0,E}$ and $\rho'\otimes\rho_{0,E}$. 

Now suppose we know after the transition happens that the initial and final system states were $\rho=\ket{\psi}\bra{\psi}$ and $\rho'=\ket{\psi'}\bra{\psi'}$ respectively, but that we are ignorant of the environment states $\phi,\phi'$. Then the correct probability distribution for $\phi,\phi'$ is a conditional one, which we can deduce with Bayes' law
\begin{align}
    \mathbb{P}(\phi,\phi'|\psi\to\psi') &= \frac{\mathbb{P}(\psi\to\psi'|\phi,\phi')\mathbb{P}(\phi)\mathbb{P}(\phi')}{\mathbb{P}(\psi\to\psi')} \\
                                        &=(\dim(\mathcal{H}_{\text{E}}))^2\frac{\abs{\mel{\Psi'}{e^{-i\mathbf{H}t/\hbar}}{\Psi}}^2}{\tr(\qty(\rho'\otimes I_{\text{E}})\,e^{-i\mathbf{H}t/\hbar}\,\qty(\rho\otimes I_{\text{E}})\,e^{i\mathbf{H}t/\hbar})} \mathbb{P}(\phi)\mathbb{P}(\phi'),
    \label{Equation: environment conditional}
\end{align}
where again $\ket{\Psi}=\ket{\psi}\otimes\ket{\phi}$ and $\ket{\Psi'}=\ket{\psi'}\otimes\ket{\phi'}$. The correlator of operators $O_i$ inserted at times $t_i$ with $i=1,\dots,m$ is also a conditional quantity. In particular, it is conditional on the initial and final environment states, and may be written
\begin{equation}
    \frac{\mel{\Psi'}{e^{-i\mathbf{H}(t-t_m)/\hbar}\,O_m\dots e^{-i\mathbf{H}(t_2-t_1)/\hbar}\,O_1\,e^{-i\mathbf{H}t_1/\hbar}}{\Psi}}{\mel{\Psi'}{e^{-i\mathbf{H}t/\hbar}}{\Psi}}.
\end{equation}
Averaging this over the conditional probability distribution~\eqref{Equation: environment conditional}, we get the expectation value of the correlator
\begin{multline}
    \expval{O_m(t_m)\dots O_1(t_1)} 
    = \sum_{\phi,\phi'}\frac{\mel{\Psi'}{e^{-i\mathbf{H}(t-t_m)/\hbar}\,O_m\dots O_1\,e^{-i\mathbf{H}t_1/\hbar}}{\Psi}\mel{\Psi}{e^{i\mathbf{H}t/\hbar}}{\Psi'}}{(\dim(\mathcal{H}_{\text{E}}))^{-2}\tr(\qty(\rho'\otimes I_{\text{E}})\,e^{-i\mathbf{H}t/\hbar}\,\qty(\rho\otimes I_{\text{E}})\,e^{i\mathbf{H}t/\hbar})} \mathbb{P}(\phi)\mathbb{P}(\phi') \\
    = \int\dd{\phi}\int\dd{\phi'} \frac{\bra{\psi'}\otimes\bra{\phi'}e^{-i\mathbf{H}(t-t_m)/\hbar}\,O_m\dots O_1\,e^{-i\mathbf{H}t_1/\hbar}\ket{\psi}\otimes\ket{\phi}\bra{\psi}\otimes\bra{\phi}e^{i\mathbf{H}t/\hbar}\ket{\psi'}\otimes\ket{\phi'}}{(\dim(\mathcal{H}_{\text{E}}))^{-2}\tr(\qty(\rho'\otimes I_{\text{E}})\,e^{-i\mathbf{H}t/\hbar}\,\qty(\rho\otimes I_{\text{E}})\,e^{i\mathbf{H}t/\hbar})}.
\end{multline}
Using~\eqref{Equation: environment resolution}, we can do this integral, and obtain
\begin{equation}
    \expval{O_m(t_m)\dots O_1(t_1)} 
    = \frac{\tr(\qty(\rho'\otimes I_{\text{E}})\,e^{-i\mathbf{H}(t-t_m)/\hbar}\,O_m\dots e^{-i\mathbf{H}(t_2-t_1)/\hbar}O_1\,e^{-i\mathbf{H}t_1/\hbar}\,\qty(\rho\otimes I_{\text{E}})\,e^{i\mathbf{H}t/\hbar})}{\tr(\qty(\rho'\otimes I_{\text{E}})\,e^{-i\mathbf{H}t/\hbar}\,\qty(\rho\otimes I_{\text{E}})\,e^{i\mathbf{H}t/\hbar})}.
\end{equation}
This is the appropriate correlator to use when we know the initial and final states of the system, but are ignorant of the environment. Note that, in the case where there are no interactions between the system and the environment, and we are only considering operators $O_i$ which act on the system, this formula can be shown to reduce to the usual one for the correlator in a closed system. However, in the presence of interactions this will not in general be true.

It is convenient to define a generating function 
\begin{equation}
    Z[\psi,\psi';J] = \tr(\qty(\rho'\otimes I_E) \, \mathbf{U}_J(t) \, \qty(\rho\otimes I_E) \, \mathbf{U}_0^\dagger(t)),
\end{equation}
where $J=J(t)$ is a time-dependent source, and $\mathbf{U}_J$ is a sourced evolution operator, defined by $\mathbf{U}_J(0)=I$ and
\begin{equation}
    i\hbar\dot{\mathbf{U}}_J\mathbf{U}_J^\dagger = \vb{H} + J \cdot O,
\end{equation}
which has the solution
\begin{equation}
    \mathbf{U}_J(T) = \mathrm{P}\exp(-\frac{i}{\hbar}\int_0^T\big(\mathbf{H}+J(t)\cdot O(t)\big)\dd{t}).
\end{equation}
Here the $\cdot$ is supposed to denote a sum over all possible operators that we want to be able to insert. Note that this generating function involves two instances of time evolution: one forwards in time with sources, and the other backwards in time but without sources. Unlike in the closed case, it is not possible in general to define a generating function with only one instance of time evolution.

Note that $\mathbf{U}_0(t) = e^{-i\mathbf{H}t/\hbar}$, so we have
\begin{equation}
    Z[\psi,\psi';0] \propto \mathbb{P}(\psi\to\psi').
\end{equation}
Also, we can compute correlators by taking appropriate derivatives of $Z$ with respect to $J$:
\begin{equation}
    \expval{O_m(t_m)\dots O_1(t_1)} = \left.\frac{(i\hbar)^m}{Z[\psi,\psi';0]}\fdv{J(t_m)}\dots\fdv{J(t_1)}Z[\psi,\psi';J]\right|_{J=0}.
\end{equation}

\subsection{Decoherence}
\label{Section: Decoherence}

Suppose the reduced state of the system is $\ket{\psi}\bra{\psi}$, and recall from the Introduction that for (according to our simple model of decoherence) after a time $\Delta t$, this state changes to
\begin{equation}
    \rho(\Delta t) \to \int \dd{X} \abs{\mel{X}{e^{-iH\Delta t/\hbar}}{\psi}}^2\,\ket{X}\bra{X}.
\end{equation}
Let us consider what this implies about the evolution of the total state in the combined system and environment. Suppose the initial combined state is
\begin{equation}
    \ket{\Psi} = \ket{\psi}\otimes \ket{\phi},
\end{equation}
where $\ket{\psi}\in\mathcal{H}$ and $\ket{\phi}\in\mathcal{H}_{\text{E}}$ are normalised. Since the total evolution of the entire system has to be unitary, $\rho(\Delta t)$ must arise as the reduced state of some pure normalised state in $\mathcal{H}\otimes\mathcal{H}_E$. In particular we can write $(\rho)\rho(\Delta t) = \tr_{\text{E}} \ket{\Psi'}\bra{\Psi'}$, where 
\begin{equation}
    \ket{\Psi'} = e^{-i\vb{H}\Delta t}\,\qty(\ket{\psi}\otimes\ket{\phi}).
\end{equation}
Using a Schmidt decomposition allows us to write this state as
\begin{equation}
    \ket{\Psi'} = \int \dd{X} f_X \ket{X}\otimes \ket{\phi_X},
\end{equation}
where 
\begin{equation}
    \abs{f_X}^2=\abs{\mel{X}{e^{-iH\Delta t/\hbar}}{\psi}}^2,
\end{equation}
and the states $\ket{\phi_X}\in\mathcal{H}_{\text{E}}$ are orthonormal with respect to the measure $\dd{X}$, i.e.\
\begin{equation}
    \braket{\phi_{X_1}}{\phi_{X_2}} = \delta(X_1,X_2).
\end{equation}
Here $\delta(X_1,X_2)$ is a Dirac delta distribution with respect to $\dd{X}$. By linearity, we must have 
\begin{equation}
    f_X = \mel{X}{e^{-iH\Delta t/\hbar}}{\psi},
\end{equation}
and $\ket{\phi_X}=Q_X\ket{\phi}$ for some operators $Q_X$ satisfying 
\begin{equation}
    \mel*{\tilde\phi}{Q_{X_1}^\dagger Q_{X_2}}{\tilde\phi} = \braket*{\tilde\phi}{\tilde\phi}\,\delta(X_1,X_2),
    \label{Equation: Q orthonormality}
\end{equation}
where $\ket*{\tilde\phi}\in\mathcal{H}_{\text{E}}$ is any environment state. So the evolution of the combined state is given by 
\begin{equation}
    \ket{\Psi}\to\ket{\Psi'}=e^{-i\vb{H}\Delta t/\hbar}\ket{\Psi},
\end{equation}
where
\begin{equation}
    e^{-i\vb{H}\Delta t/\hbar} = \int\dd{X} \ket{X}\bra{X}e^{-iH\Delta t/\hbar} \otimes Q_X.
    \label{Equation: combined evolution}
\end{equation}
The requirement of unitarity for this operator will imply some further constraints on $Q_X$, but we will not need to discuss these in detail. For evolution after a time $T=n\Delta t$ with $n$ an integer, we can use~\eqref{Equation: combined evolution} repeatedly to obtain
\begin{equation}
    e^{-i\vb{H}T/\hbar} = \int \prod_{l=1}^n\dd{X_l} \qty(\prod_{k=2}^{n}\mel{X_k}{e^{-iH\Delta t/\hbar}}{X_{k-1}}) \ket{X_n}\bra{X_1}e^{-iH\Delta t/\hbar} \otimes Q_{X_n}Q_{X_{n-1}}\dots Q_{X_1}
\end{equation}

Suppose that, during the evolution, we insert operators $O_i:\mathcal{H}\to\mathcal{H}$ at times $t_k=k \Delta t$, with $k=1,\dots,n$. We can analyse this using a sourced time evolution operator
\begin{equation}
    \vb{U}_J(t) = \qty(e^{-iJ_n\cdot O_n/\hbar}\otimes I_E)e^{-i\vb{H}\Delta t/\hbar}\dots e^{-i\vb{H}\Delta t/\hbar} \qty(e^{-iJ_1\cdot O_1/\hbar}\otimes I_E) e^{-i\vb{H}\Delta t/\hbar}.
\end{equation}
By using~\eqref{Equation: combined evolution} repeatedly, one finds that
\begin{equation}
    \vb{U}_J(t) = \int \prod_{k=1}^n \dd{X}_k U_J[X_n,X_{n-1},\dots,X_1]  \otimes Q_{X_n}Q_{X_{n-1}}\dots Q_{X_1},
    \label{Equation: sourced evolution decoherence}
\end{equation}
where
\begin{equation}
    U_J[X_n,X_{n-1},\dots,X_1] = \qty(\prod_{k=2}^{n}\mel{X_k}{e^{-iJ_k\cdot O_k/\hbar}e^{-iH\Delta t/\hbar}}{X_{k-1}}) \ket{X_n}\bra{X_1}e^{-iJ_1\cdot O_1/\hbar}e^{-iH\Delta t/\hbar}.
\end{equation}
As before, the generating function is
\begin{equation}
    Z[\psi,\psi';J] = \tr(\qty(\rho'\otimes I_E) \, \mathbf{U}_J(t) \, \qty(\rho\otimes I_E) \, \mathbf{U}_0^\dagger(t)),
\end{equation}
where $\rho=\ket{\psi}\bra{\psi}$ and $\rho'=\ket{\psi'}\bra{\psi'}$.
Actually, it will be more convenient to rescale this $Z\to Z/\dim(\mathcal{H}_E)$; this constant factor does not change the formula for computing correlators. Substituting in~\eqref{Equation: sourced evolution decoherence}, we find
\begin{multline}
    Z[\psi,\psi';J] = \int\prod_{k=1}^n\qty(\dd{X}_k\dd{\tilde{X}}_k)\tr(\rho' U_J[X_n,X_{n-1},\dots,X_1] \rho U_0[\tilde{X}_n,\tilde{X}_{n-1},\dots,\tilde{X}_1]^\dagger)\\
    \tr_{\text{E}}(Q_{X_n}Q_{X_{n-1}}\dots Q_{X_1} Q^\dagger_{\tilde{X}_1}\dots Q^\dagger_{\tilde{X}_{n-1}} Q^\dagger_{\tilde{X}_n})/(\dim(\mathcal{H}_{\text{E}}))
    \label{Equation: generating function decoherence 0}
\end{multline}
Let $\ket{i}$ be an orthonormal basis of $\mathcal{H}_{\text{E}}$. Using~\eqref{Equation: Q orthonormality}, we see that 
\begin{align}
    \tr_{\text{E}}(Q_{X_n}Q_{X_{n-1}}\dots Q_{X_1} Q^\dagger_{\tilde{X}_1}\dots Q^\dagger_{\tilde{X}_{n-1}} Q^\dagger_{\tilde{X}_n}) 
    &= \sum_i\mel{i}{Q^\dagger_{\tilde{X}_1}\dots Q^\dagger_{\tilde{X}_{n-1}} Q^\dagger_{\tilde{X}_n}Q_{X_n}Q_{X_{n-1}}\dots Q_{X_1}}{i}\\
    &= \delta(X_n,\tilde{X}_n)\sum_i\mel{i}{Q^\dagger_{\tilde{X}_1}\dots Q^\dagger_{\tilde{X}_{n-1}} Q_{X_{n-1}}\dots Q_{X_1}}{i}\\
    &= \dots = \prod_{k=1}^n\delta(X_k,\tilde{X}_k)\underbrace{\sum_i\braket{i}{i}}_{=\dim(\mathcal{H}_{\text{E}})}.
\end{align}
Substituting this into~\eqref{Equation: generating function decoherence 0} and integrating over the delta functions, we have
\begin{align}
    Z[\psi,\psi';J] &= \int\prod_{k=1}^n\dd{X}_k\tr(\rho' U_J[X_n,X_{n-1},\dots,X_1] \rho U_0[X_n,X_{n-1},\dots,X_1]^\dagger) \\
                    &= \int\prod_{k=1}^n\dd{X}_k\mel{\psi'}{U_J[X_n,X_{n-1},\dots,X_1]}{\psi}\mel{\psi}{U_0[X_n,X_{n-1},\dots,X_1]^\dagger}{\psi'}\\
                    &= \int\prod_{l=1}^n\dd{X}_l\braket{\psi'}{X_n}\mel{X_1}{e^{-iJ_1\cdot O_1/\hbar}e^{-iH\Delta t/\hbar}}{\psi}\mel{\psi}{e^{iH\Delta t/\hbar}}{X_1}\braket{X_n}{\psi'} \nonumber\\
                    &\qquad\qquad\quad\times \prod_{k=2}^n \mel{X_k}{e^{-iJ_k\cdot O_k/\hbar}e^{-iH\Delta t/\hbar}}{X_{k-1}}\mel{X_{k-1}}{e^{iH\Delta t/\hbar}}{X_k}.
\end{align}
\section{\texorpdfstring{$Y_k$}{Y\_k} in exponential form}
\label{Appendix: Y_k in exponential form}

Let $\tilde{U}_{k-1}=U_{k-1}$ and $\tilde{U}_k=e^{iH_B\Delta t/\hbar}U_k$. We will write
\begin{equation}
    Y_k=\tr(\sqrt{\rho_k}e^{-iJ_k\cdot O_k\Delta t/\hbar}e^{-iH_A\Delta t/\hbar}\sqrt{\rho_{k-1}}\tilde{U}_{k-1}^\dagger \tilde{U}_k)
    \tr(\sqrt{\rho_{k-1}}e^{iH_A\Delta t/\hbar}\sqrt{\rho_k}\tilde{U}_k^\dagger \tilde{U}_{k-1})
\end{equation}
as an exponential, assuming
\begin{align}
    \delta \tilde U_k &:= \tilde U_k - \tilde U_{k-1} = \order{\sqrt{\Delta t}},\\
    \delta \rho_k &:= \rho_k - \rho_{k-1} = \order{\sqrt{\Delta t}}.
\end{align}
We have \vspace*{-\baselineskip}
\begin{multline}
    \tr(\sqrt{\rho_k}e^{-iJ_k\cdot O_k\Delta t/\hbar}e^{-iH_A\Delta t/\hbar}\sqrt{\rho_{k-1}}\tilde U_{k-1}^\dagger \tilde U_k)
    =\tr\Big(\rho_k - \overbrace{\sqrt{\rho_k}\delta\qty(\sqrt{\rho_k})- \rho_k\delta\tilde U_k^\dagger\tilde U_k}^{=\order{\sqrt{\Delta t}}} \\
        - \underbrace{iH_k\rho_k + \sqrt{\rho_k}\delta\qty(\sqrt{\rho_k})\delta\tilde U_k^\dagger \tilde U_k}_{=\order{\Delta t}} 
        + \underbrace{i\sqrt{\rho_k}H_k\delta\qty(\sqrt{\rho_k}) + i\sqrt{\rho_k}H_k\sqrt{\rho_k}\delta \tilde U_k^\dagger \tilde U_k}_{=\order{\Delta t^{3/2}}} + \order{\Delta t^2}
    \Big)
\end{multline}
where we have defined the Hermitian operator $H_k=(H_A+J_k\cdot O_k)\Delta t/\hbar$ to slightly simplify the notation. Since $\tr(\rho_k) = 1$, we can write this as an exponential
\begin{multline}
    \tr(\sqrt{\rho_k}e^{-iJ_k\cdot O_k\Delta t/\hbar}e^{-iH_A\Delta t/\hbar}\sqrt{\rho_{k-1}}\tilde U_{k-1}^\dagger \tilde U_k) = 
    \exp\Big[ 
        \tr\Big(
            -\sqrt{\rho_k}\delta\qty(\sqrt{\rho_k})-\rho_k\delta \tilde U_k^\dagger \tilde U_k \\
            - iH_k\rho_k+\sqrt{\rho_k}\delta\qty(\sqrt{\rho_k})\delta\tilde U_k^\dagger \tilde U_k + i\sqrt{\rho_k}H_k\delta\qty(\sqrt{\rho_k})+i\sqrt{\rho_k}H_k\sqrt{\rho_k}\delta \tilde U_k^\dagger \tilde U_k
        \Big)+\\
        \tr(\sqrt{\rho_k}\delta\qty(\sqrt{\rho_k})+\rho_k\delta \tilde U_k^\dagger \tilde U_k) \tr(\sqrt{\rho_k}\delta\qty(\sqrt{\rho_k})\delta\tilde U_k^\dagger \tilde U_k-iH_k\rho_k-\frac12\qty(\sqrt{\rho_k}\delta\qty(\sqrt{\rho_k})+\rho_k\delta\tilde U_k^\dagger \tilde U_k))\\
        -\frac13\qty(\tr(\sqrt{\rho_k}\delta(\sqrt{\rho_k})+\rho_k\delta\tilde U^\dagger_k \tilde U_k))^3+ \order{\Delta t^2}
    \Big].
    \label{Equation: Y_k exp 1}
\end{multline}
Similarly, we have
\begin{multline}
    \tr(\sqrt{\rho_{k-1}}e^{iH_A\Delta t/\hbar}\sqrt{\rho_k}\tilde U_{k}^\dagger \tilde U_{k-1}) = 
    \exp\Big[ 
        \tr\Big(
            -\delta\qty(\sqrt{\rho_k})\sqrt{\rho_k}-\rho_k \tilde U_k^\dagger \delta \tilde U_k \\
            + iH^0\rho_k+\delta\qty(\sqrt{\rho_k})\sqrt{\rho_k}\tilde U_k^\dagger\delta \tilde U_k - i\delta\qty(\sqrt{\rho_k})H^0\sqrt{\rho_k}-i\sqrt{\rho_k}H^0\sqrt{\rho_k}\tilde U_k^\dagger\delta \tilde U_k
        \Big)+\\
        \tr(\delta\qty(\sqrt{\rho_k})\sqrt{\rho_k}+\rho_k\tilde U_k^\dagger\delta \tilde U_k)\tr(\delta\qty(\sqrt{\rho_k})\sqrt{\rho_k}\tilde U_k^\dagger\delta \tilde U_k+iH^0\rho_k-\frac12\qty(\delta\qty(\sqrt{\rho_k})\sqrt{\rho_k}+\rho_k\tilde U_k^\dagger\delta \tilde U_k))\\
        -\frac13\qty(\tr(\delta\qty(\sqrt{\rho_k})\sqrt{\rho_k}+\rho_k\tilde U^\dagger_k\delta \tilde U_k))^3 + \order{\Delta t^2}
    \Big],
    \label{Equation: Y_k exp 2}
\end{multline}
where $H^0 = H_A\Delta t/\hbar$. Now 
\begin{equation}
    \delta\rho_k = \rho_k-(\underbrace{\sqrt{\rho_{k-1}}}_{\mathclap{=\sqrt{\rho_k}-\delta(\sqrt{\rho_{k}})}})^2 = \sqrt{\rho_k}\delta(\sqrt{\rho_k}) + \delta(\sqrt{\rho_k})\sqrt{\rho_k} - \delta(\sqrt{\rho_k})\delta(\sqrt{\rho_k}).
\end{equation}
Using this and $\tr(\delta\rho_k)=0$ we see that
\begin{equation}
    \tr(\sqrt{\rho_k}\delta\qty(\sqrt{\rho_k})) = \frac12\tr(\sqrt{\rho_k}\delta\qty(\sqrt{\rho_k})+\delta\qty(\sqrt{\rho_k})\sqrt{\rho_k}) = \frac12\tr(\delta\qty(\sqrt{\rho_k})\delta\qty(\sqrt{\rho_k}))
\end{equation}
is actually $\order{\Delta t}$. We also have
\begin{equation}
    0 = \tilde U^\dagger_k\tilde U_k-\tilde U^\dagger_{k-1}\underbrace{\tilde U_{k-1}}_{\mathclap{=\tilde U_k-\delta \tilde U_k}} = \tilde U^\dagger_k\delta \tilde U_k + \delta \tilde U^\dagger_k \tilde U_k - \delta \tilde U_k^\dagger\delta \tilde U_k ,
\end{equation}
which we can use to show
\begin{multline}
    \sqrt{\rho_k}\delta\qty(\sqrt{\rho_k})\delta\tilde U_k^\dagger\tilde U_k + \delta\qty(\sqrt{\rho_k})\sqrt{\rho_k}\tilde U_k^\dagger\delta \tilde U_k \\
    = \frac12\qty(\sqrt{\rho_k}\delta\qty(\sqrt{\rho_k})-\delta\qty(\sqrt{\rho_k})\sqrt{\rho_k})\qty(\delta\tilde U_k^\dagger \tilde U_k-\tilde U_k^\dagger\delta \tilde U_k) + \frac12(\delta\rho_k+\delta\qty(\sqrt{\rho_k})\delta\qty(\sqrt{\rho_k}))\delta\tilde U_k^\dagger\delta \tilde U_k
\end{multline}
and
\begin{multline}
    \sqrt{\rho_k}\delta\qty(\sqrt{\rho_k})\delta\tilde U_k^\dagger \tilde U_k - \delta\qty(\sqrt{\rho_k})\sqrt{\rho_k}\tilde U_k^\dagger\delta \tilde U_k \\
    =\frac12\qty(\delta\rho_k+\delta\qty(\sqrt{\rho_k})\delta\qty(\sqrt{\rho_k}))\qty(\delta\tilde U_k^\dagger \tilde U_k-\tilde U_k^\dagger\delta \tilde U_k) + \frac12 (\sqrt{\rho_k}\delta(\sqrt{\rho_k})-\delta\qty(\sqrt{\rho_k})\sqrt{\rho_k})\delta\tilde U_k^\dagger\delta \tilde U_k.
\end{multline}
Using these formulae and other similar ones, we can compute $Y_k$ by taking the product of the two exponentials~\eqref{Equation: Y_k exp 1} and~\eqref{Equation: Y_k exp 2} above, obtaining (after a certain amount of algebra)
\begin{multline}
    Y_k = \exp\Big[\tr(-\delta\qty(\sqrt{\rho_k})\delta\qty(\sqrt{\rho_k})-\qty(\sqrt{\rho_k}\delta\qty(\sqrt{\rho_k})-\delta\qty(\sqrt{\rho_k})\sqrt{\rho_k})C_k + \tilde\rho_k C_k^2) - \tr(\tilde\rho_k C_k)^2 \\
    +\frac{i\Delta t}{\hbar}\tr(\qty(H_A+\frac12J_k\cdot O_k)\big(\delta(\sqrt{\rho_k})\sqrt{\rho_k}-\sqrt{\rho_k}\delta(\sqrt{\rho_k})-2\sqrt{\rho_k}C_k\sqrt{\rho_k}\big))\\
    + \frac{2i\Delta t}{\hbar}\tr(\rho_k\qty(H_A+\frac12 J_k\cdot O_k))\tr(\tilde\rho_kC_k)
    - \frac{i\Delta t}{\hbar} \tr(\tilde\rho_k J_k\cdot O_k) +\order{\Delta t^2}\Big],
\end{multline}
where
\begin{align}
    C_k &= \frac12\qty(\tilde U_k^\dagger\delta \tilde U_k - \delta\tilde U_k^\dagger \tilde U_k)\\
        &= \frac12\qty(U_{k-1}^\dagger e^{iH_B\Delta t/\hbar} U_k - U_k^\dagger e^{-iH_B\Delta t/\hbar}U_{k-1})
\end{align}
and
\begin{equation}
    \tilde\rho_k = \frac{\rho_k+\rho_{k-1}}2.
\end{equation}

\section{Operators in terms of modular flow}

Consider an invertible density matrix $\rho_k$ with modular Hamiltonian $K_k = -\log\rho_k$. Let the spectral decomposition of $K_k$ be given by\footnote{Technically, we could change the lower limit in this integrals to $0$, because $K_k > 0$. }
\begin{equation}
    K_k = \int_{-\infty}^\infty E \dd{\Pi_E}.
\end{equation}
Here $\dd{\Pi_E}$ is a projection valued measure, defined such that
\begin{equation}
    \Pi_{[E_1,E_2]} = \int_{E_1}^{E_2}\dd{\Pi_E}
\end{equation}
is the projector onto the space spanned by states with modular energy (i.e.\ $K_k$ eigenvalue) in the range $[E_1,E_2]$. The identity and $\rho_k$ may be written in terms of this measure as
\begin{align}
    I &= \int_{-\infty}^\infty \dd{\Pi_E}, \label{Equation: identity modular decomposition}\\
    \rho_k &= \int_{-\infty}^\infty e^{-E} \dd{\Pi_E}.
\end{align}
It is useful to note the explicit formula
\begin{equation}
    \dv{\Pi_E}{E} = \frac1{2\pi}\int_{-\infty}^\infty \dd{\alpha} e^{i\alpha E}\rho_k^{i\alpha},
    \label{Equation: modular PVM}
\end{equation}
which is just a Fourier transform.

\subsection{Infinitesimal dynamical Uhlmann holonomy \texorpdfstring{$a_k$}{a\_k}}
\label{Appendix: ak in terms of modular flow}

The operator defining infinitesimal dynamical Uhlmann holonomy is
\begin{equation}
    a_k = \int_0^\infty\dd{s} e^{-s\tilde\rho_k} \qty(\sqrt{\rho_k}\delta(\sqrt{\rho_k}) - \delta(\sqrt{\rho_k})\sqrt{\rho_k} + \frac{2i\Delta t}{\hbar}\sqrt{\rho_k}\qty(H_A+\frac12J_k\cdot O_k)\sqrt{\rho_k})e^{-s\tilde\rho_k}.
\end{equation}
Let us assume $\delta\rho_k=o(\sqrt{\Delta t})$. Then we can write 
\begin{equation}
    a_k = \int_0^\infty\dd{s} e^{-s\rho_k} \qty(\sqrt{\rho_k}\delta(\sqrt{\rho_k}) - \delta(\sqrt{\rho_k})\sqrt{\rho_k} + \frac{2i\Delta t}{\hbar}\sqrt{\rho_k}\qty(H_A+\frac12J_k\cdot O_k)\sqrt{\rho_k})e^{-s\rho_k} + o(\Delta t).
\end{equation}
We will now rewrite this in terms of modular flow. Acting with~\eqref{Equation: identity modular decomposition} on the left and right, we get
\begin{align}
    a_k &= \int_{-\infty}^\infty\dd{\Pi_E} \qty(\sqrt{\rho_k}\delta(\sqrt{\rho_k}) - \delta(\sqrt{\rho_k})\sqrt{\rho_k} + \frac{2i\Delta t}{\hbar}\sqrt{\rho_k}\qty(H_A+\frac12J_k\cdot O_k)\sqrt{\rho_k})\int_{-\infty}^\infty \dd{\Pi_{\tilde{E}}} \nonumber\\
        &\qquad\qquad\qquad\qquad\qquad\qquad\qquad\qquad\qquad\qquad\int_0^\infty\dd{s}\exp(-se^{-E})\exp(-se^{-\tilde{E}}).\\
        &= \int_{-\infty}^\infty\dd{\Pi_E} \qty(\delta(\sqrt{\rho_k})(e^{-\frac12 E}-e^{-\frac12 \tilde{E}}) + \frac{2i\Delta t}{\hbar}\qty(H_A+\frac12J_k\cdot O_k)e^{-\frac12(E+\tilde{E})})\int_{-\infty}^{\infty} \dd{\Pi_{\tilde{E}}} \frac1{e^{-E}+e^{-\tilde{E}}}.
    \label{Equation: modular a_k 1}
\end{align}
For convenience we are leaving out the $o(\Delta t)$ part. Note that for integer $n$ we have
\begin{align}
    \dd{\Pi_E}\delta(\rho_k^n)\dd{\Pi_{\tilde{E}}} &= \sum_{j=0}^{n-1} \dd{\Pi_E}\rho_k^j\delta\rho_k\rho_k^{n-1-j}\dd{\Pi_{\tilde{E}}} + o(\Delta t)\label{Equation: delta rho n start}\\
                                                   &= \dd{\Pi_E}\delta\rho_k\dd{\Pi_{\tilde{E}}} \sum_{j=0}^{n-1} e^{-jE}e^{-(n-1-j)\tilde{E}} + o(\Delta t) \\
                                                   &=\dd{\Pi_E}\delta\rho_k\dd{\Pi_{\tilde{E}}} \frac{e^{-nE} - e^{-n\tilde{E}}}{e^{-E} - e^{-\tilde{E}}} + o(\Delta t).
\end{align}
By analytic continuation of $n$, we have
\begin{align}
    \dd{\Pi_E}\delta K_k\dd{\Pi_{\tilde{E}}} &= -\left.\dv{n}\dd{\Pi_E}\delta(\rho_k^n)\dd{\Pi_{\tilde{E}}}\right|_{n=0}\\
                                              &= -\dd{\Pi_E}\delta\rho_k\dd{\Pi_{\tilde{E}}} \left.\dv{n}\frac{e^{-nE} - e^{-n\tilde{E}}}{e^{-E} - e^{-\tilde{E}}}\right|_{n=0} + o(\Delta t) \\
                                              &= \dd{\Pi_E}\delta\rho_k\dd{\Pi_{\tilde{E}}} \frac{E-\tilde{E}}{e^{-E} - e^{-\tilde{E}}} + o(\Delta t).\label{Equation: delta rho n end}
\end{align}
We can combine these to write
\begin{equation}
    \dd{\Pi_E}\delta(\sqrt{\rho_k}) \dd{\Pi_{\tilde{E}}} = \dd{\Pi_E} \delta K_k \dd{\Pi_{\tilde{E}}} \frac{e^{-\frac12 E}-e^{-\frac12 \tilde{E}}}{E-\tilde{E}} + o(\Delta t).
\end{equation}
Substituting this into~\eqref{Equation: modular a_k 1}, one finds
\begin{equation}
    a_k = \int_{-\infty}^\infty \dd{\Pi_E} \qty(\frac{1-\sech((E-\tilde{E})/2)}{E-\tilde{E}}\delta K_k + \frac{i\Delta t}{\hbar}\qty(H_A+\frac12 J_k\cdot O_k)\sech((E-\tilde{E})/2)) \int_{-\infty}^\infty \dd{\Pi_{\tilde{E}}},
\end{equation}
where again we are leaving out the $o(\Delta t)$ part.
Using now~\eqref{Equation: modular PVM}, we have
\begin{multline}
    a_k = \frac1{4\pi^2}\int_{-\infty}^\infty \dd{E}\int_{-\infty}^\infty\dd{\tilde{E}}\int_{-\infty}^\infty \dd{\alpha} \int_{-\infty}^\infty\dd{\tilde\alpha} e^{i\alpha E}e^{i\tilde\alpha \tilde{E}}\\
    \rho_k^{i\alpha} \qty(\frac{1-\sech((E-\tilde{E})/2)}{E-\tilde{E}}\delta K_k + \frac{i\Delta t}{\hbar}\qty(H_A+\frac12 J_k\cdot O_k)\sech((E-\tilde{E})/2)) \rho_k^{i\tilde\alpha}.
\end{multline}
Things simplify at this point if we change variables from $E,\tilde{E}$ to 
\begin{equation}
    x = \frac12(E+\tilde{E}),\qquad
    y = \frac12(E-\tilde{E}),
\end{equation}
so that
\begin{multline}
    a_k = \frac1{4\pi^2}\int_{-\infty}^\infty \dd{x}\int_{-\infty}^\infty\dd{y}\int_{-\infty}^\infty \dd{\alpha} \int_{-\infty}^\infty\dd{\tilde\alpha} e^{i(\alpha+\tilde\alpha)x}e^{i(\alpha-\tilde\alpha)y}\\
    \rho_k^{i\alpha} \qty(\frac{1-\sech(y)}{y}\delta K_k+ \frac{2i\Delta t}{\hbar}\qty(H_A+\frac12 J_k\cdot O_k)\sech(y)) \rho_k^{i\tilde\alpha}.
\end{multline}
The $x$ integral gives $2\pi\delta(\alpha+\tilde\alpha)$, so
\begin{equation}
    a_k = \frac1{2\pi}\int_{-\infty}^\infty\dd{y}\int_{-\infty}^\infty \dd{\alpha}e^{2i\alpha y} \rho_k^{i\alpha} \qty(\frac{1-\sech(y)}{y}\delta K_k + \frac{2i\Delta t}{\hbar}\qty(H_A+\frac12 J_k\cdot O_k)\sech(y)) \rho_k^{-i\alpha}.
    \label{Equation: modular a_k}
\end{equation}

\subsection{Bures metric \texorpdfstring{$G_k$}{G\_k}}
\label{Appendix: Gk in terms of modular flow}

The operator defining the Bures metric is
\begin{equation}
    G_k = \int \dd{s} e^{-s\rho_k} \delta\rho_k e^{-s\rho_k}.
    \label{Equation: appendix Gk}
\end{equation}
We will write this in terms of modular flow; this proceeds in much the same way as in Appendix~\ref{Appendix: ak in terms of modular flow}. In that Appendix we assumed $\delta\rho=o(\sqrt{\Delta t})$, but here we will take $\delta\rho = \order{\sqrt{\Delta t}}$. Then, by similar methods to~\eqref{Equation: delta rho n start}-\eqref{Equation: delta rho n end}, we obtain
\begin{equation}
    \dd{\Pi_E}\delta\rho_k\dd{\Pi_{\tilde{E}}} = \dd{\Pi_E}\delta K_k\dd{\Pi_{\tilde{E}}}\frac{e^{-E}-e^{-\tilde{E}}}{E-\tilde{E}} + \order{\Delta t}.
\end{equation}
From this we find that acting with~\eqref{Equation: identity modular decomposition} on the left and right of~\eqref{Equation: appendix Gk} yields
\begin{equation}
    G_k = \int_{-\infty}^{\infty} \dd{\Pi_E}\delta K_k \int_{-\infty}^\infty\dd{\Pi_{\tilde{E}}} \frac{\tanh(E-\tilde{E})}{E-\tilde{E}} + \order{\Delta t}.
\end{equation}
Now substituting in~\eqref{Equation: modular PVM} and changing variables to
\begin{equation}
    x = \frac12(E+\tilde{E}),\qquad
    y = \frac12(E-\tilde{E}),
\end{equation}
we get
\begin{equation}
    G_k = \frac1{4\pi^2}\int_{-\infty}^\infty \dd{x}\int_{-\infty}^\infty\dd{y} \int_{-\infty}^\infty\dd{\alpha}\int_{-\infty}^\infty\dd{\tilde{\alpha}} e^{i(\alpha+\tilde\alpha)x}e^{i(\alpha-\tilde\alpha)y}\frac{\tanh(y)}{y}\rho_k^{i\alpha} \delta K_k \rho_k^{i\tilde\alpha} + \order{\Delta t}.
\end{equation}
Doing the $x$ integral, we end up with
\begin{equation}
    G_k = \frac1{2\pi}\int_{-\infty}^\infty\dd{y} \int_{-\infty}^\infty\dd{\alpha} e^{2i\alpha y} \frac{\tanh(y)}{y} \rho_k^{i\alpha}\delta K_k \rho_k^{-i\alpha} + \order{\Delta t}.
\end{equation}

\printbibliography

\end{document}